\documentclass[12pt,epsf,amssymb]{article} \textheight =23 truecm
\def\lsim{\raise0.3ex\hbox{$<$\kern-0.75em\raise-1.1ex\hbox{$\sim$}}}
\def\gsim{\raise0.3ex\hbox{$>$\kern-0.75em\raise-1.1ex\hbox{$\sim$}}}
\def\noi{\noindent}  \def\bea{\begin{eqnarray}}
\def\eea{\end{eqnarray}} \def\beq{\begin{equation}}
\def\eeq{\end{equation}} 
\def\beeq{\begin{eqnarray}} \def\eeeq{\end{eqnarray}} \def\R{ {\rm R
\kern -.31cm I \kern .15cm}} \def\C{ {\rm C \kern -.15cm \vrule
width.5pt \kern .12cm}} \def\Z{ {\rm Z \kern -.27cm \angle \kern
.02cm}} \def\N{ {\rm N \kern -.26cm \vrule width.4pt \kern .10cm}}
\def\1{{\rm 1\mskip-4.5mu l} }

\usepackage{graphics} \usepackage{graphicx} \usepackage{epsfig}

\begin{document} 
\begin{center} 

{\large \bf Isgur-Wise functions and unitary representations \\ of the Lorentz group : the meson case with $j = {1 \over 2}$ light cloud} 

\par \vskip 10 truemm

 {\bf A. Le Yaouanc, L. Oliver and J.-C. Raynal}

\par \vskip 2 truemm

{\it Laboratoire de Physique Th\'eorique}\footnote{Unit\'e Mixte de
Recherche UMR 8627 - CNRS }\\    {\it Universit\'e de Paris XI,
B\^atiment 210, 91405 Orsay Cedex, France} 

\end{center}
\par \vskip 4 truemm

\begin{abstract}

We pursue the group theoretical method to study Isgur-Wise functions. We apply the general formalism, formerly applied to the baryon case $j^P = 0^+$ (for $\Lambda_b \to \Lambda_c \ell \overline{\nu}_{\ell}$), to mesons with $j^P = {1 \over 2}^-$, i.e. $\overline{B} \to D(D^{(*)})\ell\nu$. In this case, more involved from the angular momentum point of view, only the principal series of unitary representations of the Lorentz group contribute. We obtain an integral representation for the IW function $\xi(w)$ with a positive measure, recover the bounds for the slope and the curvature of $\xi(w)$ obtained from the Bjorken-Uraltsev sum rule method, and get new bounds for higher derivatives. We demonstrate also that if the lower bound for the slope is saturated, the measure is a  $\delta$-function, and $\xi(w)$ is given by an explicit elementary function. Inverting the integral formula, we obtain the measure in terms of the IW function, allowing to formulate criteria to decide if a given ansatz for the Isgur-Wise function is compatible or not with the sum rule constraints. Moreover, we have obtained an upper bound on the IW function valid for any value of $w$. We compare these theoretical constraints to a number of forms for $\xi(w)$ proposed in the literature. The "dipole" function $\xi(w) = \left({2 \over w+1} \right)^{2c}$ satisfies all constraints for $c \geq {3 \over 4}$, while the QCD Sum Rule result including condensates does not satisfy them. Special care is devoted to the Bakamjian-Thomas relativistic quark model in the heavy quark limit and to the description of the Lorentz group representation that underlies this model. Consistently, the IW function satisfies all Lorentz group criteria for any explicit form of the meson Hamiltonian at rest. 

\vskip 8 truemm
\noi LPT-Orsay-14-39 \qquad\qquad July 2014 \par 
\vskip 8 truemm

\end{abstract}

\newpage
\section{Introduction} \hspace*{\parindent} 

The heavy quark limit of QCD and, more generally, Heavy Quark Effective Theory (HQET), has aroused an enormous interest in the decade of the 1990's, starting from the formulation of Heavy Quark Symmetry by Isgur and Wise \cite{1r}.\par
Hadrons with one heavy quark such that $m_Q >> \Lambda_{QCD}$ can be thought as a bound state of a light cloud in the color source of the heavy quark. Due to its heavy mass, the latter is unaffected by the interaction with soft gluons.\par
In this approximation, the decay of a heavy hadron with four-velocity $v$ into another hadron with velocity $v'$, for example the semileptonic decay $\overline{B} \to D^{(*)}\ell \overline{\nu}_{\ell}$ or $\Lambda_b \to \Lambda_c \ell \overline{\nu}_{\ell}$, occurs just by free heavy quark decay produced by a current, and the rearrangement of the light cloud, to follow the heavy quark in the final state and constitute the final heavy hadron.\par
The dynamics is contained in the complicated light cloud, that concerns long distance QCD and is not calculable from first principles. Therefore, one needs to parametrize this physics through form factors, the IW functions.\par
The matrix element of a current between heavy hadrons containing heavy quarks $Q$ and $Q'$ can thus be factorized as follows \cite{2r}
$$< H'(v'), J'\ m'|J^{Q'Q}|H(v), J\ m >\ = \sum_{\mu,M,\mu',M'} < {1 \over 2}\ \mu', j' M'| J' m' > < {1 \over 2}\ \mu, j M| J m > $$ 
\beq
\label{1e}
\times <Q'(v'), {1 \over 2}\ \mu'|J^{Q'Q}|Q(v),{1 \over 2}\ \mu > < {\rm cloud},v',j',M'|{\rm cloud},v,j,M> 
\eeq
where $v$, $v'$ are the initial and final four-velocities, and $j$, $j'$, $M$, $M'$ are the angular momenta and corresponding projections of the initial and final light clouds, and $\mu, \mu'$ are the angular momentum projections of the heavy quark. \par
The current affects only the heavy quark, and all the soft dynamics is contained in the {\it overlap} between the initial and final light clouds $<v',j',M'|v,j,M>$, that follow the heavy quarks with the same four-velocity. This overlap is independent of the current heavy quark matrix element, and depends on the four-velocities $v$ and $v'$. The IW functions are given by these light clouds overlaps.\par
An important hypothesis has been done in writing the previous expression, namely neglecting {\it hard gluon radiative corrections}.\par
As we will make explicit below, the light cloud belongs to a Hilbert space, and transforms according to a unitary representation of the Lorentz group. Then, as we have shown in \cite{LOR-1}, the whole problem of getting rigorous constraints on the IW functions amounts to decompose unitary representations of the Lorentz group into irreducible ones. This allows to obtain for the IW functions general integral formulas in which the crucial point is that {\it  the measures are positive}.\par
In \cite{LOR-1} we did treat the case of a light cloud with angular momentum $j = 0$ in the initial and final states, as happens in the baryon semileptonic decay $\Lambda_b \to \Lambda_c \ell \overline{\nu}_{\ell}$. \par
A different but, as we will show below, equivalent method to the one of the present paper was developed in a number of articles using {\it sum rules} in the heavy quark limit, like the famous Bjorken sum rule and its generalizations \cite{4r,5r,6r,LOR-1bis,LOR-2,LOR-3}.\par
The sum rule method is completely equivalent to the method of the present paper. Indeed, starting from the sum rules one can demonstrate that an IW function, say $\xi(v.v') =\ <v'|v>$ in a simplified notation, is a function of {\it positive type}, and that one can construct a unitary representation of the Lorentz group $U(\Lambda)$ and a vector state $|\phi_0>$ representing the light cloud at rest. The IW function writes then simply (e.g. in the special case $j = 0$) : 
\beq
\label{1bise}
\xi(v.v') =\ <U(B_{v'})\phi_0|U(B_v)\phi_0>
\eeq
where $B_v$ and $B_{v'}$ are the corresponding boosts.\par
Let us now go back to previous work on the sum rule method. In the meson case $\overline{B} \to D^{(*)}\ell \overline{\nu}_{\ell}$, in the leading order of the heavy quark expansion, Bjorken sum rule (SR) \cite{4r}\cite{5r} gives the lower bound for the derivative of the IW function at zero recoil $\rho^2 = - \xi ' (1) \geq {1\over 4}$. A new SR was formulated by Uraltsev in the heavy quark limit \cite{6r} that, combined with Bjorken's, gave the much stronger lower bound $\rho^2 \geq {3 \over 4}$. A basic ingredient in deriving this bound was the consideration of the non-forward amplitude $ \overline{B}(v_i) \to D^{(n)}(v') \to  \overline{B}(v_f)$, allowing for general four-velocities $v_i$, $v_f$, $v'$.\par

In \cite{LOR-1bis,LOR-2,LOR-3} we did develop a manifestly covariant formalism within the Operator Product Expansion (OPE) and the non-forward amplitude, using the whole tower of heavy meson states \cite{2r}. We did recover Uraltsev SR plus a general class of SR that allow to bound also higher derivatives of the IW function. In particular, we found a bound on the curvature in terms of the slope $\rho^2$, namely 
\beq
\label{2e}
\xi '' (1) \geq {1 \over 5} \left [ 4 \rho^2 + 3(\rho^2)^2 \right ]
\eeq

The more powerful method of the present paper will provide a new insight on the physics of QCD in the heavy quark limit and on its Lorentz group structure.\par
As we will see below, we obtain an integral formula for the Isgur-Wise function in terms of a {\it positive measure}. We will see that we recover the bound (\ref{2e}) and that this systematic method allows to find bounds for higher derivatives.\par
We can invert this integral formula and obtain the measure corresponding to any given ansatz for the IW function and we obtain thus a powerful criterium to decide if this ansatz is consistent with the Lorentz group approach or, equivalently, with the generalized Bjorken-Uraltsev sum rules.
The method exposed in this paper allows to decide if a given model for the IW function is consistent or not with general principles of QCD in the heavy quark limit.\par

The purpose of the present paper is purely theoretical. In HQET, e.g. in $b \to c$ transitions, one can take the heavy quark limit for both initial and final quarks while keeping finite the mass ratio $r = m_b/m_c$. Varying the ratio $r$ one can in principle attain any value for the variable $w$ within the range $1 \leq w \leq {1+r^2 \over 2r}$, and our theoretical constraints on IW functions are then valid for any value of $w$.\par
Of course, this is quite different from the physical range at finite masses, namely $1 \leq w \leq 1.4$ GeV. To perform an analysis at finite mass would ask not only to implement the theoretical constraints on the IW function obtained in the present work. One would need to perform a serious phenomenological discussion and to include $1/m_Q$ corrections, radiative corrections within the effective theory HQET, and make use of the Wilson coefficients to make the matching with the true QCD, as has been done for the curvature of the IW function (\ref{2e}) by M. Dorsten \cite{DORSTEN}. This whole program is outside the intention of the present work, that only deals with rigorous constraints on the shape of the IW function. 

The outline of the paper is as follows. Sections 2 and 3 recall necessary generalities and details on the present Lorentz group approach to IW functions, following closely ref. \cite{LOR-1}. In Section 4 we particularize the method exposed in detail in \cite{LOR-1} to the present meson case, making explicit the needed unitary representations of the Lorentz group. In Section 5 we compute the irreducible IW functions in the case $j = {1 \over 2}$ and give an integral formula expressing the IW function in terms of the latter and a positive mesure. In Section 6 we use this integral formula to get a polynomial expression for the derivatives of the IW function, and in Section 7 we obtain lower bounds on its derivatives. Section 8 is devoted to obtain the inversion of the integral formula for the IW function. In Section 9 we find an {\it upper bound} on the IW function. In Section 10 we apply the inverted integral formula to study consistency tests of a number of models of the IW function proposed in the literature. The Bakamjian-Thomas relativistic quark model in the heavy quark limit and the description of the Lorentz group representation that underlies this model is studied in detail in Section 11. In Section 12 we discuss the theoretical and phenomenological relevance of our results, and we conclude.

\section{The Lorentz group and the heavy quark limit of QCD} \hspace*{\parindent}
In the heavy mass limit, the states of a heavy hadron $H$ containing a heavy quark $Q$ is described as follows \cite{2r}, as we can see from (\ref{1e}) :
\beq
\label{3e}
|H(v),\mu,M>\ = |Q(v),\mu> \otimes\ |v,j,M>
\eeq
where there is factorization into the heavy quark state factor $|Q(v),\mu>$ and a light cloud component $|v,j,M>$. The velocity $v$ of the heavy hadron H is the same as the velocity of the heavy quark $Q$, and is unquantized. The heavy quark $Q$ state depends only on a spin $\mu = \pm {1\over2}$ quantum number, and so belongs to a 2-dimensional Hilbert space. The light component is the complicated thing, but it does not depend on the spin state $\mu$ of the heavy quark $Q$, nor on its mass, and this gives rise to the symmetries of the heavy quark theory.\par
As advanced in the Introduction, the matrix element of a heavy-heavy current $J$ (acting only on the heavy quark) writes

$$< H'(v'),\mu',M'|J|H(v),\mu,M>\ =\ <Q'(v'),\mu'|J|Q(v),\mu>$$
\beq
\label{4e}
\times <v',j',M'|v,j,M> 
\eeq
and the IW functions are defined as the coefficients, depending only on $v.v'$, in the expansion of the unknown scalar products $<v',j',M'|v,j,M>$ into independent scalars constructed from $v$, $v'$ and the polarization tensors describing the spin states of the light components.\par
Now, the crucial point in the present work is that {\it the states of the light components make up a Hilbert space in which acts a unitary representation of the Lorentz group}. In fact, this is more or less implicitly stated, and used in the literature \cite{2r}.\par

\subsection{Physical picture of a heavy quark} \hspace*{\parindent}
To see the point more clearly, let us go into the physical picture which is at the basis of (\ref{3e}). Considering first a heavy hadron {\it at rest}, with velocity

\beq
\label{4bise}
v_0 = (1, \vec{0})
\eeq
its light component is submitted to the interactions between the light particles, light quarks, light antiquarks and gluons, and to the external chromo-electric field generated by the heavy quarks at rest. This chromo-electric field does not depend on the spin $\mu$ of the heavy quark nor on its mass. We shall then have a complete orthonormal system of energy eigenstates $|v_0,j,M,\alpha>$ of the light component, where $j$ and $M$ are the angular momentum quantum numbers and $\alpha$ designs other quantum numbers (like the radial excitation number),  
\beq
\label{5e}
<v_0,j',M',\alpha'|v_0,j,M,\alpha>\ = \delta_{j,j'} \delta_{M,M'} \delta_{\alpha,\alpha'} 
\eeq

Now, for a heavy hadron moving with a velocity $v$, the only thing which changes for the light component is that the external chromo-electric field generated by the heavy quark at rest is replaced by the external chromo-electromagnetic field generated by the heavy quark moving with velocity $v$. Neither the Hilbert space describing the possible states of the light component, nor the interactions between the light particles, are changed. We shall then have {\it a new complete orthonormal system} of energy eigenstates $|v,j,M,\alpha>$, in the same Hilbert space. Then, because the colour fields generated by a heavy quark for different velocities are related by Lorentz transformations, we may expect that the energy eigenstates of the light component will, for various velocities, be themselves related by Lorentz transformations acting in their Hilbert space.\par

\subsection{Lorentz representation from covariant overlaps} \hspace*{\parindent}
Let us now show that such a representation of the Lorentz group does in fact underly the work of ref. \cite{2r}. The description of spin states by polarization tensors is used.\par

For half-integer spin $j$, in which we are interested in the present paper, the polarization tensor becomes a Rarita-Schwinger tensor-spinor $\epsilon^{\mu_1,...\mu_{j-1/2}}_\alpha$ subject to the constraints of symmetry, transversality and tracelessness
\beq
\label{6e}
v_{\mu_1} \epsilon^{\mu_1,...\mu_{j-1/2}}_\alpha = 0     \qquad\qquad\qquad	 g_{\mu_1\mu_2}\ \epsilon^{\mu_1,\mu_2...\mu_{j-1/2}}_\alpha = 0
\eeq

\noindent and
\beq
\label{7e}
({/\hskip - 2 truemm v}-1)_{\alpha\beta} \epsilon^{\mu_1,...,\mu_{j-1/2}}_\beta  = 0    \qquad\qquad	 (\gamma_{\mu_1})_{\alpha\beta} \epsilon^{\mu_1,...,\mu_{j-1/2}}_\beta = 0
\eeq

Then a scalar product $<v',j',\epsilon'|v,j,\epsilon>$ is a covariant function of the vectors $v$ and $v'$ and of the tensors (or tensor-spinors) $\epsilon'^*$ and $\epsilon$, bilinear with respect to $\epsilon'^*$ and $\epsilon$, and the IW functions, functions of the scalar $v.v'$, are introduced accordingly.\par
The covariance property of the scalar products is explicitly expressed by the equality
\beq
\label{8e}
<\Lambda v',j',\Lambda \epsilon'|\Lambda v,j,\Lambda \epsilon>\ =\ <v',j',\epsilon'|v,j,\epsilon> 
\eeq 
valid for any Lorentz transformation $\Lambda$, with the transformation of a tensor-spinor given by
\beq
\label{10e}
(\Lambda \epsilon)^{\mu_1,...,\mu_{j-1/2}}_\alpha = \Lambda^{\mu_1}_{\nu_1} ...  \Lambda^{\mu_{j-1/2}}_{\nu_{j-1/2}} D(\Lambda)_{\alpha\beta}\ \epsilon^{\nu_1,...,\nu_{j-1/2}}_\beta
\eeq

Then, let us {\it define} the operator $U(\Lambda)$, in the space of the light cloud states, by
\beq
\label{11e}
U(\Lambda)|v,j,\epsilon>\ = |\Lambda v,j,\Lambda\epsilon> 
\eeq 
where here $v$ is a fixed, arbitrarily chosen velocity. Eq. (\ref{8e}) implies that $U(\Lambda)$ is a {\it unitary operator}, as demonstrated in \cite{LOR-1}.

\subsection{From a Lorentz representation to Isgur-Wise functions} \hspace*{\parindent}

A unitary representation of the Lorentz group emerges thus from the usual treatment of heavy hadrons in the heavy quark theory. For the present purpose, we need to go in the opposite way, namely, to show how, starting from a unitary representation of the Lorentz group, the usual treatment of heavy hadrons and the introduction of the IW functions emerges. What follows is not restricted to the $j = {1 \over 2} $ case, but concerns any IW function.\par
So, let us consider some unitary representation $\Lambda \to U(\Lambda)$ of the Lorentz group, or more precisely of the group $SL(2,C)$, in a Hilbert space $\mathcal{H}$, and we have to identify states in $\mathcal{H}$, depending on a velocity $v$. As explained in \cite{LOR-1}, we have in $\mathcal{H}$ an additional structure, namely the energy operator of the light component {\it for a heavy quark at rest}, with $v_0=(1,0,0,0)$. Since this energy operator is invariant under rotations, we consider the subgroup $SU(2)$ of $SL(2,C)$. By restriction, the representation in $\mathcal{H}$ of $SL(2,C)$ gives a representation $R \to U(R)$ of $SU(2)$, and its decomposition into irreducible representations of $SU(2)$ is needed. We then have the eigenstates $|v_0,j,M>$ of the energy operator, classified by the angular momentum number $j$ of the irreducible representations of $SU(2)$, and {\it associated with the rest velocity} $v_0$, since their physical meaning is to describe the energy eigenstates of the light component for a heavy quark at rest.\par
We need now to express the states $|v,j,\epsilon>$ in terms of the states $|v_0,j,M>$. We begin with $v = v_0$. For fixed $j$ and $\alpha$, the states $|v_0,j,M>$ constitute, for $-j \leq M \leq j$, a standard basis of a representation $j$ of $SU(2)$ :  
\beq
\label{13e}
U(R)\ |v_0,j,M>\ = \sum_{M'}\ D^j_{M',M}(R)\ |v_0,j,M'> 
\eeq
where the rotation matrix elements $D^j_{M',M}$ are defined by
\beq
\label{40e}
D^j_{M',M} =\ <j,M'|U_j(R)|j,M>   \qquad\qquad  R \in SU(2)
\eeq

On the other hand, the states $|v_0,j,\epsilon>$ constitute, when $\epsilon$ goes over all polarization tensors (or tensor-spinors), the whole space of a representation $j$ of $SU(2)$. As emphasized in \cite{LOR-1}, this representation of $SU(2)$ in the space of 3-tensors (or 3-tensor-spinors) is not irreducible, but contains an irreducible subspace of spin $j$, which is precisely the polarization 3-tensor (or 3-tensor-spinor) space selected by the other constraints (\ref{6e}) and (\ref{7e}) for velocity $v_0$.\par
We may then introduce a standard basis $\epsilon^{(M)}$, $-j \leq M \leq j$, for the $SU(2)$ representation of spin $j$ in the space of polarization 3-tensors (or 3-tensor-spinors). As demonstrated in \cite{LOR-1}, the states $|v,j,\epsilon>$ are given by
\beq
\label{19e}
|v,j,\epsilon>\ = \sum_{M} (\Lambda^{-1}\epsilon)_M\ U(\Lambda) |v_0,j,M>
\eeq
for any $\Lambda$ such that $\Lambda v_0 = v$, with $v_0$ given by (\ref{4bise}), and $(\Lambda^{-1}\epsilon)_M$ is the component of the velocity $v_0$ polarization tensor $\Lambda^{-1}\epsilon$ in the stadard basis.
 
Equation (\ref{19e}) is {\it our final result} here, defining, in the Hilbert space $\mathcal{H}$ of a unitary representation of $SL(2,C)$, the states $|v,j,\epsilon>$ which transform as (\ref{11e}) and whose scalar products define the IW functions, in terms of $|v_0,j,M>$ which occur as $SU(2)$ multiplets in the restriction to $SU(2)$ of the $SL(2,C)$ representation. And these states $|v,j,\epsilon>$ defined by (\ref{19e}) do indeed transform as (\ref{11e}). \par

\section{Decomposition into irreducible representations and integral formula for the IW function} \hspace*{\parindent}

In the case of a compact group (as $SU(2)$), any unitary representation can be written as a direct sum of irreducible ones. In the present case of $SL(2,C)$ (a non-compact group), the more general notion of a direct integral is required \cite{10r}. 
Let us denote by $X$ the set of irreducible unitary representations of $SL(2,C)$, by $\mathcal{H}_\chi$ the Hilbert space of a representation $\chi \in X$, and by $U_\chi(\Lambda)$ the unitary operator acting in $\mathcal{H}_\chi$ which corresponds to any $\Lambda \in SL(2,C)$. Then, for any unitary representation of $SL(2,C)$, the Hilbert space $\mathcal{H}$ can be written in the form
\beq
\label{22e}
\mathcal{H} = \int_{X}^{\oplus} \oplus_{n_\chi} \mathcal{H}_\chi\ d\mu(\chi)
\eeq
where $\mu$ is an arbitrary {\it positive} measure on the set $X$ and $n_\chi$ is a function on $X$ with $\geq 1$ integer values or possibly $\infty$. Explicitly, an element $\psi \in \mathcal{H}$ is a function
\beq
\label{23e}
\psi : \chi \in X \to \psi_\chi = (\psi_{1,\chi},... ,\psi_{n_\chi,\chi})  \in \oplus_{n_\chi} \mathcal{H}_\chi
\eeq
which assigns to each $\chi \in X$ an element $\psi_\chi \in \oplus_{n_\chi} \mathcal{H}_\chi$, and which is $\mu$-measurable and square $\mu$-integrable. The scalar product in $\mathcal{H}$ is given by :
\beq
\label{24e}
<\psi '|\psi>\ = \int_{X} <\psi_\chi '|\psi_\chi> d\mu(\chi)
\eeq
and the operator $U(\Lambda)$ of the representation in the space $\mathcal{H}$ is given by :
\beq
\label{25e}
\left (U(\Lambda\right )\psi)_{k,\chi} = U_\chi(\Lambda)\psi_{k,\chi}
\eeq

Let us see now the consequences for the IW functions. For simplicity, we take here the case of a spinor ($j = {1 \over 2} $) light component. For the hadron at rest, the light component will be described by {\it some} element $\psi_{1 \over 2} \in \mathcal{H}$ {\it which is spinor} for the subgroup $SU(2)$ of $SL(2,C)$. Then, according to the transformation law (\ref{25e}), requiring that $\psi_{1 \over 2}$ is a spinor under rotations is the same as requiring that $\psi_{{1 \over 2},k,\chi}$ is a spinor under rotations for all $\chi$'s and all $k = 1,... , n_\chi$. More generally, the decomposition of the irreducible representations of $SL(2,C)$ into irreducible representations of $SU(2)$ is known (see Section below). Since $SU(2)$ is compact, the decomposition is by a direct sum, and therefore each $\mathcal{H}_\chi$ admits an orthonormal basis adapted to $SU(2)$. Moreover, it turns out that each representation $j$ of $SU(2)$ appears with multiplicity 0 or 1. Then, there is a subset $X_0 \subset X$ of irreducible representations of $SL(2,C)$ containing a non-zero $SU(2)$ spinor subspace and, for $\chi \in X_0$, there is a unique (up to a phase) normalized $SU(2)$ scalar element in $\mathcal{H}_\chi$, which we denote $\phi_{{1 \over 2},\chi}$. Each scalar element in $\mathcal{H}_\chi$ is then proportional to $\phi_{{1 \over 2},\chi}$. So, one has
\beq
\label{26e}
\psi_{{1 \over 2},\chi} = (c_{1,\chi}\ \phi_{{1 \over 2},\chi},... ,c_{n_\chi,\chi}\ \phi_{{1 \over 2},\chi})
\eeq
with some coefficients $c_{1,\chi},... ,c_{n_\chi,\chi}$.
From the scalar product  (\ref{24e}) in $\mathcal{H}$, one sees that the normalization $<\psi_{1 \over 2}|\psi_{1 \over 2}>\ = 1$ of the light component amounts to

\beq
\label{27e}
\int_{X_0}\ \sum_{k=1}^{n_\chi}\ |c_{k,\chi}|^2\ d\mu(\chi) = 1
\eeq

\section{Lorentz group irreducible unitary representations and their decomposition under rotations} 

\subsection{Explicit form of the principal series of irreducible unitary representations of the Lorentz group} \hspace*{\parindent}

We have described in \cite{LOR-1} an explicit form of the irreducible unitary representations of $SL(2,C)$. Their set $X$ is divided into three sets, the set $X_p$ of representations of the principal series, the set $X_s$ of representations of the supplementary series, and the one-element set $X_t$ made up of the trivial representation \cite{11r}.\par
Actually, for the $j = {1 \over 2}$ case, only the principal series is relevant. For the moment, let us however consider the principal series, leaving $j$ completely general. \par
A representation $\chi = (n,\rho)$ in the principal series is labelled by an integer $n \in Z$ and a real number $\rho \in R$. Actually, the representations $(n,\rho)$ and $(-n,-\rho)$ (as given below) turn out to be equivalent so that, in order to have each representation only once, $n$ and $\rho$ will be restricted as follows \cite{11r} :
$$n = 0 \qquad\qquad\qquad \rho \geq 0$$
\beq
\label{33e}
n > 0 \qquad\qquad\qquad \rho \in R 
\eeq

\noindent Notice that we keep the standard notation $\rho$ used in mathematical books to label the irreducible Lorentz group representations. This parameter should not be confused with the also standard notation in HQET for the slope of the IW function $\rho^2$.

The Hilbert space $\mathcal{H}_{n,\rho}$ is made up of functions of a complex variable $z$ with the standard scalar product
\beq
\label{34e}
<\phi'|\phi>\ = \int \overline{\phi'(z)}\ \phi(z)\ d^2z
\eeq
with the measure $d^2z$ in the complex plane being simply $d^2z = d(Re z)d(Im z)$. So $\mathcal{H}_{n,\rho} = L^2(C,d^2z)$.\par
The unitary operator $U_{n,\rho}(\Lambda)$ is given by :
\beq
\label{35e}
\left(U_{n,\rho}(\Lambda)\phi \right)\!(z) =  \left({{\alpha-\gamma z} \over {|\alpha-\gamma z|}}\right)^n |\alpha-\gamma z|^{2i\rho-2}\ \phi\!\left({\delta z-\beta} \over {\alpha-\gamma z}\right)
\eeq
where $\alpha$, $\beta$, $\gamma$, $\delta$ are complex matrix elements of $\Lambda \in SL(2,C)$ :

\beq
\label{36e}
\Lambda = \left( \begin{array}{cc} \alpha & \beta \\ \gamma & \delta \end{array} \right) \qquad\qquad\qquad \alpha \delta - \beta \gamma = 1
\eeq

\subsection{Decomposition under the rotation group} \hspace*{\parindent}
Next we need the decomposition of the restriction to the subgroup $SU(2)$ of each irreducible unitary representation of $SL(2,C)$.\par
Since $SU(2)$ is compact, the decomposition is by a direct sum so that, for each representation $\chi \in X$ we have an {\it orthonormal basis} $\phi^\chi_{j,M}$ of $\mathcal{H}_\chi$ adapted to $SU(2)$. Having in mind the usual notation for the spin of the light component of a heavy hadron, here we denote by $j$ the spin of an irreducible representation of $SU(2)$. It turns out \cite{11r} that each representation $j$ of $SU(2)$ appears in $\chi$ with multiplicity 0 or 1, so that $\phi^\chi_{j,M}$ needs no more indices, and that the values taken by $j$ are part of the integer and half-integer numbers. For fixed $j$, the functions $\phi^\chi_{j,M}$, $-j \leq M \leq j$ are choosen as a standard basis of the representation $j$ of $SU(2)$. \par
It turns out \cite{11r} that the functions $\phi^\chi_{j,M}(z)$ are expressed in terms of the rotation matrix elements $D^j_{M',M}$ defined by (\ref{40e}). A matrix $R \in SU(2)$ being of the form

\beq
\label{37e}
R = \left( \begin{array}{cc} a & b \\ -\overline{b} & \overline{a} \end{array} \right) \qquad\qquad\qquad |a|^2+|b|^2 = 1
\eeq
we shall also consider $D^j_{M',M}$ as a function of $a$ and $b$, satisfying $|a|^2+|b|^2 = 1$.\par

We can now give explicit formulae for the orthonormal basis $\phi^\chi_{j,M}$ of $\mathcal{H}_\chi$.\par

The spins $j$ which appear in a representation $\chi = (n,\rho)$ are \cite{LOR-1} :
\beq
\label{38-1e}
\qquad {\rm all\ integers} \qquad \qquad \qquad \ \ j \geq {n \over 2} \qquad \ \ \  {\rm for} \qquad n \qquad {\rm even} \qquad
\eeq
\beq
\label{38-2e}
{\rm all\ half-integers} \qquad \qquad j \geq {n \over 2} \qquad \ \  {\rm for} \qquad n \qquad {\rm odd}
\eeq

\noindent Such a spin appears with multiplicity 1.\par The basis functions $\phi^{n,\rho}_{j,M}(z)$ are given by the expression \cite{LOR-1}
\beq
\label{45e}
\phi^{n,\rho}_{j,M}(z) =  {\sqrt{2j+1} \over \sqrt{\pi}}\ (1+|z|^2)^{i\rho-1} D^j_{n/2,M}\! \left( {1 \over \sqrt{1+|z|^2}}, - {z \over \sqrt{1+|z|^2}} \right)	
\eeq
or, using an explicit formula for $D^j_{n/2,M}$ :
$$\phi^{n,\rho}_{j,M}(z) = {\sqrt{2j+1} \over \sqrt{\pi}}\ (-1)^{n/2-M}\sqrt{{(j-n/2)!(j+n/2)!} \over {(j-M)!(j+M)!}}\ (1+|z|^2)^{i\rho-j-1}$$ 
\beq
\label{46e}
\sum_{k}\ (-1)^k  \left( \begin{array}{c} j+M \\ k \end{array} \right) \left( \begin{array}{c} j-M \\ j-n/2-k \end{array} \right) z^{n/2-M+k}\ \overline{z}^k 
\eeq
where the range for $k$ can be limited to $0 \leq k \leq{j-n/2}$ due to the binomial factors.\par

\section{Irreducible Isgur-Wise functions for $j = {1 \over 2}$} \hspace*{\parindent}

For $j = {1 \over 2}$, one has a fixed value for $n$
\beq
\label{47e}
j = {1 \over 2} \qquad \Rightarrow \qquad n = 1, \qquad \rho \in R
\eeq

\noindent and we are thus in the case (\ref{38-2e}).\par
Deleting from now on  the fixed indices $j = {1 \over 2}$ and $n = 1$, and particularizing the explicit formula (\ref{46e}) to this case, the non-vanishing functions (\ref{46e}) read :   
\beq
\label{48-1e}
\phi^{\rho}_{+ {1 \over 2}}(z) = \sqrt{2 \over \pi} \left(1 + |z|^2 \right)^{i\rho - {3 \over 2}} \qquad
\eeq
\beq
\label{48-2e}
\phi^{\rho}_{- {1 \over 2}}(z) = -\sqrt{2 \over \pi}\ z\left(1 + |z|^2 \right)^{i\rho - {3 \over 2}} \ 
\eeq

Let us now particularize the $SL(2,C)$ matrix (\ref{36e}) to a boost in the $z$ direction :
\beq
\label{49e}
\Lambda_\tau = \left( \begin{array}{cc} e^{\tau \over 2} & 0 \\ 0 & e^{- {\tau \over 2}} \end{array} \right) \qquad \qquad \qquad w = \cosh(\tau)
\eeq

\noindent and, following the $j = 0$ case studied at length in \cite{LOR-1}, let us consider the following objects
\beq
\label{50-1e}
\xi^\rho_{+ {1 \over 2}, + {1 \over 2}}(w) =\ <\phi^\rho_{+ {1 \over 2}}|U^\rho(\Lambda_\tau)\phi^\rho_{+ {1 \over 2}}>
\eeq
\beq
\label{50-2e}
\xi^\rho_{- {1 \over 2}, - {1 \over 2}}(w) =\ <\phi^\rho_{- {1 \over 2}}|U^\rho(\Lambda_\tau)\phi^\rho_{- {1 \over 2}}>
\eeq

From the transformation law (\ref{35e}) and the explicit forms (\ref{48-1e}),(\ref{48-2e}), one gets :
\beq
\label{51-1e}
\left(U^\rho(\Lambda_\tau)\phi^\rho_{+ {1 \over 2}}\right)(z) = \sqrt{2 \over \pi}\ e^{(i\rho-1)\tau}\left(1 + e^{-2\tau}|z|^2 \right)^{i\rho - {3 \over 2}}
\eeq
\beq
\label{51-2e}
\left(U^\rho(\Lambda_\tau)\phi^\rho_{- {1 \over 2}}\right)(z) = - \sqrt{2 \over \pi}\ e^{(i\rho-1)\tau} e^{-\tau} z \left(1 + e^{-2\tau}|z|^2 \right)^{i\rho - {3 \over 2}}
\eeq

\noindent and therefore, from these expressions and (\ref{50-1e}),(\ref{50-2e}), one obtains :
\beq
\label{52-1e}
\xi^\rho_{+ {1 \over 2}, + {1 \over 2}}(w) = {2 \over \pi} \int \left(1 + |z|^2 \right)^{-i\rho - {3 \over 2}} e^{(i\rho-1)\tau} \left(1 + e^{-2\tau}|z|^2 \right)^{i\rho - {3 \over 2}} d^2z \qquad \qquad 
\eeq
\beq
\label{52-2e}
\xi^\rho_{- {1 \over 2}, - {1 \over 2}}(w) = {2 \over \pi} \int e^{-\tau} |z|^2 \left(1 + |z|^2 \right)^{-i\rho - {3 \over 2}} e^{(i\rho-1)\tau} \left(1 + e^{-2\tau}|z|^2 \right)^{i\rho - {3 \over 2}} d^2z \ \ \ 
\eeq

We must now extract the Lorentz invariant Isgur-Wise function $\xi(w)$. To do that, we must decompose into invariants the matrix elements (\ref{52-1e}),(\ref{52-2e}) using the spin ${1 \over 2}$ spinors of the light cloud $u_{\pm {1 \over 2}}$. We have not introduced parity in our formalism. Therefore, we will have the following decomposition :
\beq
\label{53-1e}
\xi^\rho_{+ {1 \over 2}, + {1 \over 2}}(w) = \left(\overline{u}_{+ {1 \over 2}}(v')u_{+ {1 \over 2}}(v) \right) \xi^\rho(w) + \left(\overline{u}_{+ {1 \over 2}}(v')\gamma_5u_{+ {1 \over 2}}(v) \right) \tau^\rho(w)
\eeq
\beq
\label{53-2e}
\xi^\rho_{- {1 \over 2}, - {1 \over 2}}(w) = \left(\overline{u}_{- {1 \over 2}}(v')u_{- {1 \over 2}}(v) \right) \xi^\rho(w) + \left(\overline{u}_{- {1 \over 2}}(v')\gamma_5u_{- {1 \over 2}}(v) \right) \tau^\rho(w)
\eeq

\noindent where $\xi^\rho(w)$ is an irreducible ${1 \over 2}^- \to {1 \over 2}^-$ elastic IW function, labelled by the index $\rho$, and $\tau^\rho(w)$ is a function corresponding to the flip of parity ${1 \over 2}^- \to {1 \over 2}^+$.\par
The notation for the function $\tau^\rho(w)$ has to be distinguished from the one for the boost parameter $\tau$ introduced in (\ref{49e}).\par
Let us now compute the spinor bilinears of relations (\ref{53-1e}),(\ref{53-2e}). From the expression 
\beq
\label{54e}
u_{\pm {1 \over 2}}(v) = \sqrt{v^0+1 \over 2}\left( \begin{array}{c} \chi_{\pm {1 \over 2}} \\ {{\bf \sigma}.{\bf v} \over v^0+1} \chi_{\pm {1 \over 2}} \end{array} \right) \qquad \qquad \qquad \overline{u}_{\pm {1 \over 2}}(v) u_{\pm {1 \over 2}}(v) = 1
\eeq

\noindent one gets
\beq
\label{54-1e}
\overline{u}_{+ {1 \over 2}}(v')u_{+ {1 \over 2}}(v) = \overline{u}_{- {1 \over 2}}(v')u_{- {1 \over 2}}(v) = \sqrt{w+1 \over 2}
\eeq
\beq
\label{54-2e}
\overline{u}_{+ {1 \over 2}}(v')\gamma_5 u_{+ {1 \over 2}}(v) = - \overline{u}_{- {1 \over 2}}(v') \gamma_5 u_{- {1 \over 2}}(v) = {1 \over \sqrt{2}}\ \sqrt{w-1 \over w+1}
\eeq

\noindent and therefore we obtain
\beq
\label{55-1e}
\xi^\rho(w) = \sqrt{2 \over w+1}\ {1 \over 2}\ \left[\xi^\rho_{+ {1 \over 2}, + {1 \over 2}}(w) + \xi^\rho_{- {1 \over 2}, - {1 \over 2}}(w) \right]
\eeq
\beq
\label{55-2e}
\tau^\rho(w) = \sqrt{2}\ \sqrt{w+1 \over w-1}\ {1 \over 2}\ \left[\xi^\rho_{+ {1 \over 2}, + {1 \over 2}}(w) - \xi^\rho_{- {1 \over 2}, - {1 \over 2}}(w) \right]
\eeq

\noindent and from expressions (\ref{52-1e})(\ref{52-2e}) one gets finally :
\beq
\label{56-1e}
\xi^\rho(w) = {1 \over 1+\cosh(\tau)}{1 \over \sinh(\tau)}\ {1 \over 2} \left[{e^{(i\rho-{1\over2})\tau}-e^{-(i\rho-{1\over2})\tau} \over i\rho-{1\over2}} + {e^{(i\rho+{1\over2})\tau}-e^{-(i\rho+{1\over2})\tau} \over i\rho+{1\over2}}\right]
\eeq

\noindent or
\beq
\label{56-2e}
\xi^\rho(w) = {1 \over 1+\cosh(\tau)}{1 \over \sinh(\tau)}\ {4 \over 4\rho^2+1}\left[\sinh\left({\tau \over 2} \right)\cos(\rho\tau)+2\rho\cosh\left({\tau \over 2}\right)\sin(\rho\tau)\right]
\eeq

This is the expression for the elastic ${1 \over 2}^- \to {1 \over 2}^-$ irreducible IW functions we were looking for, parametrized by the real parameter $\rho$, that satisfies 
\beq
\label{56-3e}
\xi^\rho(1) = 1
\eeq

Like in the case $j = 0$, analized in great detail in \cite{LOR-1}, the elastic ${1 \over 2}^- \to {1 \over 2}^-$ IW function $\xi(w)$ will be given by the integral over a positive measure $d\nu(\rho)$ :
\beq
\label{57-1e}
\xi(w) = \int_{]-\infty,\infty[} \xi^\rho(w)\ d\nu(\rho)
\eeq
\noindent where the measure is normalized acording to
\beq
\label{57-2e}
\int_{]-\infty,\infty[} d\nu(\rho) = 1
\eeq

\noindent Notice that the range $]-\infty,\infty[$ for the parameter $\rho$ that labels the irreducible representations follows from the fact that in the $j = {1 \over 2}$ case one has $n = 1$ and $\rho \in R$, eq. (\ref{47e}). Notice also that the IW irreducible function (\ref{56-2e}) is even in $\rho$, $\xi^\rho(w) = \xi^{-\rho}(w)$. This seems to contradict the non-equivalence of the irreducible representations labelled by $\rho$ and $-\rho$, but this can be resolved by considering the Lorentz plus parity group.\par
The irreducible IW functions (\ref{56-2e}), parametrized by some value of $\rho = \rho_0$, are legitimate IW functions since the corresponding measure is given by a delta function,
\beq
\label{57-3e}
d\nu(\rho) = \delta(\rho-\rho_0)\ d\rho
\eeq

In the case of the irreducible representation $\rho_0 = 0$ one finds
\beq
\label{57-4e}
\xi^0(w) = {4 \sinh\left({\tau \over 2}\right) \over (1+\cosh(\tau))\sinh(\tau)} = \left({2 \over 1+w}\right)^{3 \over 2}
\eeq

\noindent that saturates the lower bound for the slope $-\xi'(1) \geq {3 \over 4}$. This is the so-called BPS limit of the IW function, considered previously using different theoretical arguments \cite{BPSU, BPSJLOR}.

\section{Integral formula for the IW function $\xi(w)$ and polynomial expression for its derivatives} 

From the normalization of the norm (\ref{57-2e}) and the normalization of the irreducible IW functions (\ref{56-3e}) one gets the correct value of the IW function at zero recoil 
\beq
\label{59e}
\xi(1) = 1
\eeq

The integral formula (\ref{56-2e})and (\ref{57-1e}) writes, explicitly,
$$\xi(w) = {1 \over 1+\cosh(\tau)}{1 \over \sinh(\tau)}$$
\beq
\label{60-1e}
\times  \int_{]-\infty,\infty[}{4 \over 4\rho^2+1} \left[\sinh\left({\tau \over 2} \right)\cos(\rho\tau)+2\rho\cosh\left({\tau \over 2}\right)\sin(\rho\tau)\right]\ d\nu(\rho)
\eeq
\noindent from which one can find the following polynomial expression for its derivatives :
\beq
\label{60-2e}
\xi^{(n)}(1) = (-1)^n {1 \over 2^{2n}(2n+1)!!} \prod_{i=1}^{n} \left< \left[ (2i+1)^2+4\rho^2 \right] \right> \qquad \qquad (n \ge 1)
\eeq

\noindent where the mean value in (\ref{60-2e}) is defined as follows :
\beq
\label{60-3e}
\left< f(\rho) \right> = \int_{]-\infty,\infty[} f(\rho)\ d\nu(\rho)  
\eeq

Formula (\ref{60-2e}) can be demonstrated along the same lines as the corresponding one in the baryon case (Appendix D of ref. \cite{LOR-1}) by using the following integral representation of the irreducible IW function (\ref{56-1e}) or (\ref{56-2e}) :
\beq
\label{60-4e}
\xi^\rho(w) = {2 \over \rho^2 + {1 \over 4}} {\cosh(\pi \rho) \over \pi} \int_0^\infty x^{-i\rho+{1 \over 2}} {1+x \over (1+2wx+x^2)^2}\ dx
\eeq

\section{Bounds on the derivatives of the IW function} 

\subsection{Lower bounds on the derivatives}

Bounds on the successive derivatives of the IW function are important. Indeed, the extrapolation at zero recoil to obtain $\mid V_{cb} \mid$ from the semileptonic exclusive data is sensitive to high derivatives (curvature and third derivative, at least) because the data points are more precise at large recoil than at low recoil \cite{LOR-2}.\par

From the expression (\ref{60-2e}) one gets immediately the lower bounds on the derivatives
\beq
\label{61e}
(-1)^n \xi^{(n)}(1) \ge {(2n+1)!! \over 2^{2n}}
\eeq

\noindent obtained in \cite{MLOPR-2}, that reduces for the slope and the curvature to the bounds
\beq
\label{62e}
- \xi'(1) \ge {3 \over 4}\ , \qquad \qquad \qquad \xi''(1) \ge {15 \over 16}
\eeq

\subsection{Improved bounds on the derivatives}

To get improved bounds on the derivatives we must, like in \cite{LOR-1}, express the derivatives in terms of moments of the {\it positive} variable $\rho^2$, that we can read from (\ref{60-2e}). Calling the moments : 
\beq
\label{63e}
\mu_n =\ <\rho^{2n}>\ \ge\ 0 \qquad \qquad (n \ge 0)
\eeq

\noindent one gets the successive derivatives in terms of moments :
$$\xi(1) = \mu_0 = 1$$
$$\xi'(1) = - \left({3 \over 4} + {1 \over 3} \mu_1 \right)$$
$$\xi''(1) = {15 \over 16} + {17 \over 30} \mu_1 + {1 \over 15} \mu_2$$
\beq
\label{64e}
\xi^{(3)}(1) = - \left({105 \over 64} + {1891 \over 1680} \mu_1 + {83 \over 420} \mu_2 + {1 \over 105} \mu_3 \right)
\eeq
$$\xi^{(4)}(1) = {945 \over 256} + {4561 \over 1680} \mu_1 + {4307 \over 7560} \mu_2 + {41 \over 945} \mu_3 + {1 \over 945} \mu_4 $$

\noindent etc.

Notice that the lowest bounds (\ref{61e}) and (\ref{62e}) are found in the limit $\mu_n = 0$.

The equations (\ref{64e}) can be solved step by step, and the moment $\mu_n$ is expressed as a combination of the derivatives $\xi(1)$, $\xi'(1)$,... $\xi^{(n)}(1)$ :
$$\mu_0 = \xi(1) = 1$$
$$\mu_1 = - {3 \over 4} \left[ 3+4\xi'(1) \right]$$
$$\mu_2 = {3 \over 16} \left[ 27+136\xi'(1)+80\xi''(1) \right]$$
\beq
\label{65e}
\mu_3 = - {3 \over 64} \left[ 243+3724\xi'(1)+6640\xi''(1)+2240\xi^{(3)}(1) \right]
\eeq
$$\mu_4 = {3 \over 256}  \left[ 2187 + 96016\xi'(1) + 399840\xi''(1)+367360\xi^{(3)}(1) +80640\xi^{(4)}(1) \right]$$
\noindent etc.

Since $\rho^2$ is a positive variable, one can obtain improved bounds on the derivatives from the following set of constraints. For any $n \geq 0$, one has \cite{LOR-1}
\beq
\label{79e}
det\left[(\mu_{i+j})_{0\leq i,j \leq n}\right] \geq 0
\eeq
\beq
\label{80e}
det\left[(\mu_{i+j+1})_{0\leq i,j \leq n}\right] \geq 0
\eeq

Since each moment $\mu_k$ is a combination of the derivatives $\xi(1)$, $\xi'(1)$,... $\xi^{(k)}(1)$, the constraints on the moments translate into constraints on the derivatives.\par
We shall treat here in detail only the constraints on $\mu_1$, $\mu_2$, $\mu_3$, which are given respectively by (\ref{80e}) ($n = 0$), (\ref{79e}) ($n = 1$), (\ref{80e}) ($n = 1$) :
\beq
\label{81e}
\mu_1 \geq 0 \qquad\qquad\qquad\qquad\qquad\qquad\qquad\qquad\qquad\qquad\qquad
\eeq
\beq
\label{82e}
det \left( \begin{array}{cc}
1&\mu_1\\
\mu_1&\mu_2\\
\end{array} \right) = \mu_2 - \mu_1^2 \geq 0 \qquad\qquad\qquad\qquad\qquad\qquad
\eeq

\beq
\label{83e}
det \left( \begin{array}{cc}
\mu_1&\mu_2\\
\mu_2&\mu_3\\
\end{array} \right) = \mu_1 \mu_3 - \mu_2^2 \geq 0 \qquad\qquad\qquad\qquad\qquad\qquad
\eeq

\beq
\label{84e}
det \left( \begin{array}{ccc}
1&\mu_1&\mu_2\\
\mu_1&\mu_2&\mu_3\\
\mu_2&\mu_3&\mu_4\\
\end{array} \right) = (\mu_2 - \mu_1^2)\mu_4 - (\mu_3^2 - 2 \mu_1 \mu_2 \mu_3 + \mu_2^3) \geq 0
\eeq

\noindent etc.\par
Clearly, each moment $\mu_k$ is bounded from below, and the lower bound is given by (\ref{79e}) for $k$ even and by (\ref{80e}) for $k$ odd in terms of the lower moments. So (\ref{81e})-(\ref{84e}) give :
\beq
\label{85e}
\mu_1 \geq 0 \qquad\qquad\qquad\qquad\qquad\qquad\qquad\qquad\qquad
\eeq

\beq
\label{86e}
\mu_2 \geq \mu_1^2 \qquad\qquad\qquad\qquad\qquad\qquad\qquad\qquad\qquad
\eeq

\beq
\label{87e}
\mu_3 \geq {\mu_2^2 \over \mu_1} \qquad\qquad\qquad\qquad\qquad\qquad\qquad\qquad\qquad
\eeq

\beq
\label{87bise}
\mu_4 \geq {-\mu_2^3 + 2 \mu_1 \mu_2 \mu_3 - \mu_3^2 \over \mu_1^2 - \mu_2} \qquad\qquad\qquad \qquad\qquad
\eeq

\noindent etc.

The constraints (\ref{85e})-(\ref{87e}) imply, respectively, in terms of the derivatives :

\beq
\label{88e}
-\xi'(1) \geq {3 \over 4} \qquad\qquad\qquad\qquad\qquad\qquad\qquad\qquad\qquad\qquad\qquad\qquad\
\eeq
\beq
\label{89e}
\xi''(1) \geq {1 \over 5}\left[-4\xi'(1)+3\xi'(1)^2\right] \qquad\qquad\qquad\qquad\qquad\qquad\qquad \ \ \ 
\eeq
\beq
\label{90e}
-\xi^{(3)}(1) \geq {5 \over 28}{-12\xi'(1)+9\xi'(1)^2-39\xi''(1)-12\xi'(1)\xi''(1)+16\xi''(1)^2 \over -3-4\xi'(1)}
\eeq

\noindent from (\ref{87bise}) we find a lower bound on $\xi^{(4)}(1)$, etc. \par
We see that we recover the bounds obtained using the SR method.\par

The lower bound of the third derivative (\ref{90e}) is apparently singular for the lower bound (\ref{88e}) of the first derivative $-\xi'(1)$. However, using the lower bound (\ref{89e}) to eliminate $\xi''(1)$ we find the less restrictive lower bound
\beq
\label{91e}
-\xi^{(3)}(1) \geq {-\xi'(1)[10-3\xi'(1)][4-3\xi'(1)] \over 35}
\eeq

\section{Inversion of the integral representation of the Isgur-Wise function}

Let us now show that the integral formula for the IW function (\ref{57-1e}) can be inverted, giving the positive measure $d\nu(\rho)$ in terms of the IW function $\xi(w)$. This will allow to formulate criteria to test the validity of a given phenomenological ansatz of $\xi(w)$.\par
Let us define
\beq
\label{91-1e}
\widehat{\xi}(\tau) = (\cosh(\tau)+1)\sinh(\tau)\xi(\cosh(\tau))
\eeq

\noindent and similarly for the irreducible IW function
\beq
\label{91-2e}
\widehat{\xi}^\rho(\tau) = (\cosh(\tau)+1)\sinh(\tau)\xi_\rho(\cosh(\tau))
\eeq 

The integral formula (\ref{57-1e}) then writes
\beq
\label{92e}
\widehat{\xi}(\tau) = \int \widehat{\xi}^\rho(\tau)d\nu(\rho)
\eeq 

It is convenient to use the form (\ref{56-1e}) for the irreducible IW function. One finds, for its derivative, the simple formula :\par
\beq
\label{93-1e}
{d \over d\tau}\ \widehat{\xi}^\rho(\tau) = 2 \cos(\rho\tau)\cosh\left({\tau \over 2}\right)
\eeq

We now assume that the general measure $d\nu(\rho)$ is even, i.e. like the measure $d\rho$, without loss of generality because $\xi^\rho(w)$ is even in $\rho$. This means that $\int f(\rho) d\nu(\rho) = \int f(-\rho) d\nu(\rho)$ for any function $f(\rho)$.\par  

Defining the function
\beq
\label{93-2e}
\eta(\tau) = {1 \over 2\cosh\left({\tau \over 2}\right)} {d \over d\tau}\ \widehat{\xi}(\tau)
\eeq  

\noindent one sees, from (\ref{93-1e}), that the integral formula (\ref{57-1e}) reads
\beq
\label{94e}
\eta(\tau) = \int_{]-\infty,\infty[} \cos(\rho \tau) d\nu(\rho) = \int_{]-\infty,\infty[} e^{-i\rho \tau} d\nu(\rho)
\eeq

Computing the Fourier transform

\beq
\label{95e}
{\tilde \eta}(\rho) = {1 \over 2\pi}\int_{-\infty}^{+\infty} e^{i\tau\rho}\ d\tau\ \eta(\tau) = \int_{]-\infty,\infty[} \delta(\rho-\rho')\ d\nu(\rho')
\eeq 

\noindent and defining the function
\beq
\label{96e}
\mu(\rho) = {d\nu(\rho) \over d\rho}
\eeq

\noindent one finds
\beq
\label{97e}
{\tilde \eta}(\rho) = \mu(\rho) 
\eeq

The function (\ref{96e}) is even  
\beq
\label{97-1e}
\mu(\rho) = \mu(-\rho) 
\eeq

\noindent and one finally finds 
\beq
\label{98-1e}
d\nu(\rho) = {\tilde \eta}(\rho) d\rho 
\eeq

\noindent or
\beq
\label{98-2e}
{d\nu(\rho) \over  d\rho} = {1 \over 2\pi}\int_{-\infty}^{+\infty} e^{i\tau\rho}\ d\tau\ {1 \over 2\cosh\left({\tau \over 2}\right)} {d \over d\tau} \left[(\cosh(\tau)+1)\sinh(\tau)\xi(\cosh(\tau))\right] 
\eeq

This completes the inversion of the integral representation. Equation (\ref{98-2e}) is the master formula expressing the measure in terms of a given ansatz for the Isgur-Wise function.\par
We can now apply this formula to check if a given phenomenological formula for the IW function $\xi(w)$ satisfies the constraint that the corresponding measure $d\nu(\rho)$ must be positive. This provides a powerful consistency test for any proposed ansatz. Notice also that (\ref{98-1e}) and (\ref{57-2e}) imply, as a necessary condition on $\xi(w)$, that the function $\eta(\rho)$, defined by (\ref{91-1e}) and (\ref{93-2e}) in terms of $\xi(w)$, {\it must be bounded} by 1.

\section{An upper bound on the Isgur-Wise function}

Also, an upper bound on the whole IW function $\xi(w)$ can be obtained from the integral formula obtained above.\par
Defining the function
\beq
\label{upper-1e}
\eta^\rho(\tau) = {1 \over 2\cosh\left({\tau \over 2}\right)} {d \over d\tau}\ \widehat{\xi}^\rho(\tau)
\eeq  

\noindent we have obtained, from (\ref{93-1e}) and (\ref{93-2e}), 
\beq
\label{upper-2e}
\eta^\rho(\tau) =\cos(\rho \tau)
\eeq

\noindent and it follows
\beq
\label{upper-3e}
-1 \leq \eta^\rho(\tau) \leq 1
\eeq

\noindent that writes
\beq
\label{upper-4e}
-2 \cosh\left({\tau \over 2}\right) \leq {d \over d\tau}\ \widehat{\xi}^\rho(\tau) \leq 2 \cosh\left({\tau \over 2}\right)
\eeq

Integrating this inequality from $0$, one gets :
\beq
\label{upper-5e}
- 4 \sinh\left({\tau \over 2}\right) \leq \widehat{\xi}^\rho(\tau) \leq 4 \sinh\left({\tau \over 2}\right)
\eeq

\noindent and since
\beq
\label{upper-6e}
\widehat{\xi}^0(\tau) = 4 \sinh\left({\tau \over 2}\right)
\eeq

\noindent one finds the inequalities 
\beq
\label{upper-7e}
- \widehat{\xi}^0(\tau) \leq \widehat{\xi}^\rho(\tau) \leq \widehat{\xi}^0(\tau)
\eeq

\noindent and simplifying common factors dependent on $\tau$ :
 \beq
\label{upper-8e}
- \xi^0(\tau) \leq \xi(\tau) \leq \xi^0(\tau)
\eeq
Since $\xi^0(\tau)$ is given by the expression (\ref{57-4e}), we finally obtain
\beq
\label{upper-9e}
| \xi(w) |\ \leq  \left({2 \over 1+w}\right)^{3 \over 2}
\eeq

\noindent This inequality is a strong result because it holds for any value of $w$.

\section{Consistency tests for any ansatz of the IW function : phenomenological applications}

In this Section we examine a number of phenomenological formulas proposed in the past in the literature.\par
We will compare these ansatze with the theoretical criteria formulated in the two preceding Sections, concerning respectively the lower bounds on derivatives at zero recoil (Section 7), the upper bound obtained for the whole IW function (Section 8), and the inversion of the integral formula for the IW function, and check of the positivity of the measure (\ref{98-2e}). For the bounds on the derivatives, we will limit the test up to the third derivative, formulas (\ref{88e})-(\ref{90e}), although the method can be generalized to any higher derivative in a straightforward way.\par
We must underline that the satisfaction of the bounds on the derivatives and of the upper bound on the whole IW function are {\it necessary conditions}, while the criterium of the positivity of the measure is a {\it necessary and sufficient condition} to establish if a given ansatz of the IW function satisfies the Lorentz group criteria of the present paper.

To illustrate the methods exposed in this paper, we use a number of proposed phenomenological models for the IW funtion. Some of these functions could happen to be rather close numerically in the physical range at finite mass $1 \leq w \leq w_{max} \simeq 1.4$ GeV. However, as underlined in the introduction, our purpose is mainly theoretical and has the interest of giving theoretical criteria as to whether a given model for the IW function satisfies or does not satisfy the general principles of QCD in the heavy quark limit.

\subsection{The exponential ansatz}

\beq
\label{99-1e}
\xi(w) = \exp\left[-c(w-1)\right] 
\eeq

This form corresponds to the non-relativistic limit for the light quark with the harmonic oscillator potential \cite{JLOR}.

\subsubsection{Bounds on the derivatives}

The bound for the slope (\ref{88e}) is satisfied for $c \geq {3 \over 4}$, the bound for the second derivative (\ref{89e}) is satisfied for $c \geq 2$, while the bound for the third derivative (\ref{90e}) is violated for any value of c.\par
Therefore, this phenomenological ansatz on the IW function is invalid on the theoretical grounds of Section 7.\par

\subsubsection{Upper bound on the IW function}

The exponential ansatz (\ref{99-1e}) satisfies nevertheless the upper bound (\ref{upper-6e}) $\xi(w) \leq  \left({2 \over 1+w}\right)^{3 \over 2}$.

\subsubsection{Positivity of the measure}

Let us now examine the criterium based on the positivity of the measure.\par 
One needs to compute
\beq
\label{99-2e}
\eta(\tau) = {1 \over c} \left(- {d^2 \over d\tau^2} + {1 \over 4} \right) \cosh\left({\tau \over 2}\right)\exp\left[-c(\cosh(\tau)-1)\right]
\eeq

\noindent The function $\eta(\tau)$ is bounded for any value of $c$ (Fig. 1).

\vskip 3 truemm

\includegraphics[scale=1.]{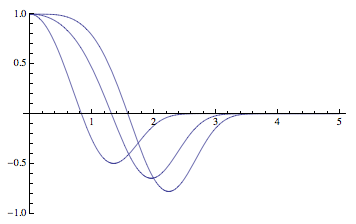}\\
Fig. 1. $\eta(\tau)$ (\ref{93-2e}) for the exponential ansatz $c = 3/4, 1, 2$ (higher to lower curves). 

\vskip 3 truemm

The Fourier transform of this function gives, from (\ref{98-1e}), 
\beq
\label{99-3e}
d\nu(\rho) = {e^c \over 2\pi} {1 \over c} \left( \rho^2 + {1 \over 4} \right) \left[ K_{i\rho + {1 \over 2}}(\rho) + K_{-i\rho + {1 \over 2}}(\rho) \right] d\rho
\eeq

This function is not positive for any value of $c$, as we illustrate in Fig. 2.

\includegraphics[scale=1.]{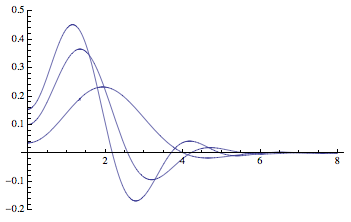}\\
Fig. 2. ${d\nu(\rho) \over d\rho}$ (\ref{98-2e}) for the exponential ansatz, for $c = 3/4, 1, 2$ (higher to lower curves). 

\vskip 3 truemm

Therefore, the exponential ansatz for the IW function violates the consistency criteria exposed in Sections 7 and 8.

\subsection{The "dipole"}

The following shape has been proposed in the literature (see for example \cite{NRSX,13bisr})
\beq
\label{100-1e}
\xi(w) = \left({2 \over 1+w}\right)^{2c} 
\eeq

\subsubsection{Bounds on the derivatives}

The bound for the slope (\ref{88e}) is satisfied for $c \geq {3 \over 4}$, while the bounds for the second derivative (\ref{89e}) and third derivative (\ref{90e}) are also satisfied for $c \geq 3/4$. The "dipole" ansatz is thus valid for any value of $c$.\par

\subsubsection{Upper bound on the IW function}

Of course, the "dipole" satisfies the upper bound (\ref{upper-6e}) $\xi(w) \leq  \left({2 \over 1+w}\right)^{3 \over 2}$ for $c \geq 3/4$.

\subsubsection{Positivity of the measure}

Let us verify this result in all generality computing the measure (\ref{98-2e}).\par 
One needs first to compute
\beq
\label{100-2e}
\eta(\tau) = -4(c-1) \left[\cosh\left({\tau \over 2}\right) \right]^{-4c+3} + (4c-3) \left[\cosh\left({\tau \over 2}\right)\right]^{-4c+1}
\eeq

Since one needs the function $\eta(\tau)$ to be bounded, the parameter $c$ must satisfy
\beq
\label{100-3e}
c \geq {3 \over 4}
\eeq

We realize that in the particular case
\beq
\label{100-4e}
c = {3 \over 4} \qquad \ \ \to \qquad \ \ \eta(\tau) = 1 \qquad \ \ \to \qquad \ \ d\nu(\rho) = \delta(\rho)\ d\rho
\eeq

Therefore, one gets in this case a delta-function for the measure, that is positive and corresponds to the explicit formula (\ref{57-4e}) for the IW function in the BPS limit given above.\par
On the other hand, one sees from the lower bound (\ref{100-3e}), that the so-called Meson Dominance IW proposal \cite{K}
\beq
\label{100-5e}
\xi_{MD}(w) = {2 \over w+1}
\eeq

\noindent does not satisfy our general constraints because in this case $c = {1 \over 2}$.

For $c > {3 \over 4}$ one obtains a function $\eta(\tau)$ that is bounded, as we can see from formula (\ref{100-6e}) and Fig. 3.\par

\vskip 3 truemm

\includegraphics[scale=1.]{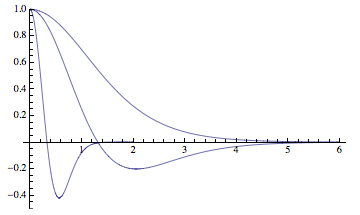}\\
Fig. 3. $\eta(\tau)$ (\ref{93-2e}) for the "dipole" ansatz $c = 1., 1.5, 2.$ (from higher to lower curves). 

\vskip 3 truemm

Computing its Fourier transform (\ref{98-2e}) one gets the measure  
\beq
\label{100-6e}
d\nu(\rho) = {2^{4c-1} \over 2\pi}\ (4c-3) \left( \rho^2 + {1 \over 4} \right){\Gamma\left(i\rho+2c-{3 \over 2}\right)\Gamma\left(-i\rho+2c-{3 \over 2}\right) \over \Gamma\left(4c-1\right)} \ d\rho
\eeq

\noindent that is positive (Fig. 4).

\includegraphics[scale=1.]{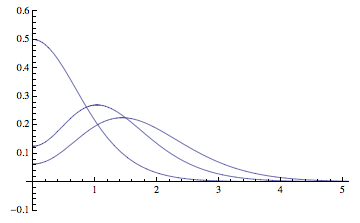}\\
Fig. 4. ${d\nu(\rho) \over d\rho}$ for the "dipole" ansatz for $c = 1., 1.5, 2.$ (from higher to lower curves). 

\vskip 3 truemm

In conclusion, from (\ref{100-4e}) and (\ref{100-6e}), we see that the measure $d\nu(\rho)$ for the "dipole" ansatz is positive for $c \geq {3 \over 4}$. Therefore, the "dipole" form satisfies all the consistency criteria.

\subsection{Kiselev's ansatz}

V. Kiselev \cite{K} proposed the following shape

\beq
\label{101-1e}
\xi(w) = \sqrt{2 \over w^2+1}\exp\left(-\beta {w^2-1 \over w^2+1}\right) 
\eeq

\noindent where $\beta = {m^2_{sp} \over \omega^2}$ and the slope is given by $\xi'(1) = - {1 \over 2} - \beta$.\par

\subsubsection{Bounds on the derivatives}

The bound for the slope (\ref{88e}) is satisfied for $\beta \geq {1 \over 4}$, the bound for the second derivative (\ref{89e}) is satisfied for $\beta \geq 0.4$, while the bound for the third derivative (\ref{90e}) is satisfied for $\beta \geq 1.5$.
One can suspect that bounds for higher derivatives will only be satisfied for higher values of $\beta$.\par

\subsubsection{Upper bound on the IW function}

Kiselev formula (\ref{101-1e}) does not satisfy the upper bound (\ref{upper-9e}) $\xi(w) \leq  \left({2 \over 1+w}\right)^{3 \over 2}$ for any value of $\beta$ because, as we can see, in the limit of large $w$ it becomes $\sqrt{2 \over w^2+1}\ e^{-\beta}$.

\subsubsection{Positivity of the measure}

One finds for the function $\eta(\tau)$ (\ref{93-2e})
\beq
\label{101-2e}
\eta(\tau) = {1 \over 2} \cosh\left({\tau \over 2}\right) \exp\left[- 2\beta {\sinh^2(\tau) \over 3+\cosh(2\tau)}\right] \left[{1 \over {3 + \cosh(2\tau)}}\right]^{5 \over 2} \times
\eeq
$$[-38+2\left(53+8\beta\right)\cosh(\tau)-24\cosh(2\tau)+21\cosh(3\tau)-16\beta\cosh(3\tau)-2\cosh(4\tau)+\cosh(5\tau)]$$ 

Independently of any value of the parameter $\beta = {m^2_{sp} \over \omega^2}$, this function is not bounded since it blows up for $\tau \to \pm \infty$. Therefore the ansatz (\ref{101-1e}) for the IW function does not satisfy the general Lorentz group criteria formulated in the present paper.

\subsection{BSW formula for the IW function}

Using the relativistic oscillator wave functions of Bauer, Stech and Wirbel \cite{BSW} one finds the IW function \cite{NR}

\beq
\label{102-1e}
\xi_{BSW}(w) = \sqrt{{2 \over w+1}} {1 \over w} \exp\left(- c^2 {w-1 \over {2w}} \right) {F\left(c \sqrt{{w+1 \over 2w}} \right) \over F(c)} 
\eeq

\noindent with $c = {\alpha \over \omega}$ in the notation of \cite{BSW}, and
\beq
\label{102-2e}
F(x) = \int_{-x}^{+\infty} dz (z+x) e^{-z^2} = {1 \over 2} \left[e^{-x^2} + \sqrt{\pi} x (1 + \rm{erf}(x))\right]
\eeq

As we will see, this ansatz for the IW function allows to illustrate in detail the consistency criteria developped in this paper.\par

\subsubsection{Bounds on the derivatives}

First, the bound for the slope (\ref{88e}) is satisfied for any value of $c$ (for $c = 0$, the slope is $-\xi'_{BSW}(1) = {5 \over 4}$), while the bounds for the second derivative (\ref{89e}) and third derivative (\ref{90e}) is satisfied for any value of $c$. Up to this third derivative, the BSW ansatz seems thus valid for any value of $c$.\par

\subsubsection{Upper bound on the IW function}

The BSW formula (\ref{102-1e}) satisfies the upper bound (\ref{upper-6e}) $\xi(w) \leq  \left({2 \over 1+w}\right)^{3 \over 2}$ for any value of the parameter $c$.

\subsubsection{Positivity of the measure}

We will check now that this is true in all generality, for any derivative, using the criterium of positivity of the measure $d\nu(\rho)$ (\ref{98-2e}).\par

Computing the function $\eta(\tau)$ (\ref{93-2e}) for the BSW ansatz (\ref{102-1e}) one finds, numerically, the functions $\eta_{BSW}(\tau)$ of Fig. 5. 

\vskip 3 truemm

\includegraphics[scale=1.]{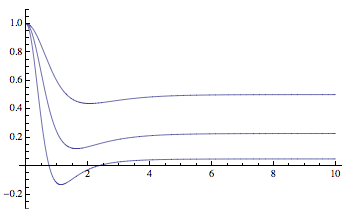}\\
Fig. 5. The function $\eta_{BSW}(\tau)$ (\ref{93-2e}) for $c = 0, 1, 2$ (from higher to lower curves). 

\vskip 4 truemm

We observe that for $\tau \to \infty$, the function $\eta(\tau)$ tends to a constant, that is found to be
\beq
\label{102-3e}
\eta^{(\infty)} = \lim_{\tau \to \infty} \eta(\tau) = {2 + c\sqrt{2\pi}\exp\left({c^2 \over 2}\right)\left[1+\rm{erf}\left({c \over \sqrt{2}}\right) \right] \over 4 + 4c\sqrt{\pi}\exp(c^2)\left[1+\rm{erf}(c) \right]}
\eeq

Since the function $\eta(\tau)$ tends to a constant, its Fourier fransform, that gives the measure (\ref{98-1e}), will contain a $\delta$-function. Substracting the constant (\ref{102-3e}), we define a new function
\beq
\label{102-4e}
\eta^{(0)}_{BSW}(\tau) = \eta_{BSW}(\tau)-\eta^{(\infty)}
\eeq

We plot this function in Fig. 6 for some values of c, and observe that it is bounded.

\vskip 10 truemm

\includegraphics[scale=1.]{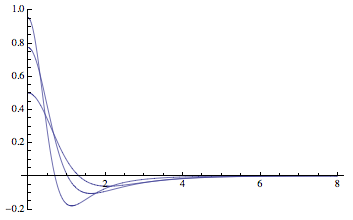}\\
Fig. 6. The function $\eta^{(0)}_{BSW}(\tau)$ (\ref{102-4e}) .   

\vskip 10 truemm

Defining, like in (\ref{95e}), its Fourier transform by 
\beq
\label{102-5e}
{{\tilde \eta}^{(0)}_{BSW}}(\rho) = {1 \over 2\pi}\int_{-\infty}^{+\infty} e^{i\tau\rho}\ \eta^{(0)}_{BSW}(\tau)d\tau 
\eeq

\noindent we obtain the functions of Fig. 7. 

\vskip 10 truemm

\includegraphics[scale=1.]{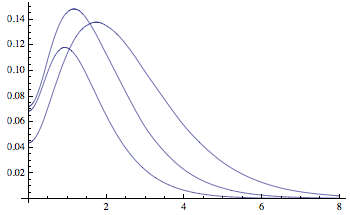}\\
Fig. 7. Fourier transform $\tilde{\eta}^{(0)}_{BSW}(\rho)$ of the function $\eta^{(0)}_{BSW}(\tau)$.

\vskip 4 truemm

Finally, the total measure will be given by
\beq
\label{102-6e}
d\nu_{BSW}(\rho) = {{\tilde \eta}^{(0)}_{BSW}}(\rho) d\rho + \eta^{(\infty)} \delta(\rho)d\rho 
\eeq

\noindent with ${{\tilde \eta}^{(0)}_{BSW}}(\rho)$ given in Fig. 3 and the constant $\eta^{(\infty)}$ by (\ref{102-3e}).

The conclusion is that the BSW ansatz for the IW function is consistent. It satisfies the theoretical criteria since both pieces of the measure (\ref{102-6e}) ${{\tilde \eta}^{(0)}_{BSW}}(\rho) d\rho$ and $\eta^{(\infty)} \delta(\rho)d\rho$ are positive. Therefore, the BSW ansatz is thus valid for any value of $c$. However, this conclusion is only based on numerical calculation. We do not have by now a complete proof.

\subsection{Relativistic harmonic oscillator}

The following shape follows from a relativistic quark model with harmonic oscillator wave function \cite{NRSX}
\beq
\label{103-1e}
\xi(w) = {2 \over w+1} \exp\left(- \beta {w-1 \over w+1}\right) 
\eeq

\noindent where the parameter $\beta$ is related to the slope by $\beta = -2\xi'(1)-1$.\par 

\subsubsection{Bounds on the derivatives}

We find that the first and second derivatives satisfy the bounds of Section 7 for $\beta \geq {1 \over 2}$, while the third derivative satisfies the constraint (\ref{90e}) for $\beta > 0.73$.\par 

\subsubsection{Upper bound on the IW function}

The formula (\ref{103-1e}) does not satisfy the upper bound (\ref{upper-6e}) $\xi(w) \leq  \left({2 \over 1+w}\right)^{3 \over 2}$ for any value of $\beta$ because, as we can see, in the limit of large $w$ it becomes a pole.

\subsubsection{Positivity of the measure}

This ansatz for the IW function does not satisfy the general consistency criterium of Section 8 for any value of $\beta$.\par 
One finds for the function $\eta(\tau)$ (\ref{93-2e})
\beq
\label{103-2e}
\eta(\tau) = {1\over 4\cosh^3\left({\tau \over 2}\right)} \exp\left[-\beta \tanh^2\left({\tau \over 2}\right)\right]
\eeq
$$\times \left[1 + 4\beta + (2-4\beta)\cosh(\tau) + \cosh(2\tau) \right]$$ 
\noindent This function is unbounded for any value of the parameter $\beta$, and therefore the proposal (\ref{103-1e}) does not satisfy the general criteria.\par
This means that bounds on some higher derivatives, as can be generalized following Section 7, are not satisfied for any given value of $\beta$.

\subsection{The IW function in the QCD Sum Rules approach}

The QCD Sum Rule approach yields the following result for the IW function, {\it switching off the hard gluon radiative corrections} \cite{MNEUBERT,NRSX} :

\beq
\label{104-1e}
\xi_{QCDSR}(w) = {{3 \over {8 \pi^2}} \left({2 \over {w+1}}\right)^2 I \left(\sigma(w) {\delta \over \Lambda} \right) + C(\Lambda,w)  \over {3 \over {8 \pi^2}} I \left({\delta \over \Lambda} \right) + C(\Lambda,1)}
\eeq

\noindent where
\beq
\label{104-2e}
C(\Lambda,w) = \left\{- {<\overline{q} q> \over \Lambda^3} \left[1 - {1 \over 6} (w-1) {4\lambda^2 \over \Lambda^2} \right] + \left({w-1 \over w+1} \right) {<\alpha_sGG> \over 48\pi\Lambda} \right\} \exp \left[- {(w+1) \over 2} {4\lambda^2 \over \Lambda^2} \right]
\eeq

\noindent and  
\beq
\label{104-3e}
I(x) = \int_{0}^{x} dy y^2 e^{-y} = 2 - (x^2+2x+2)\ e^{-x}
\eeq

\noindent On the other hand, the function $\sigma(w)$ satisfies $\sigma(1) = 1$ and is bounded by 
\beq
\label{104-4e}
{1 \over 2} (x+1-\sqrt{x^2-1}) \leq \sigma(x) \leq 1
\eeq

Let us now compute the functions $\eta(\tau)$ (\ref{94e}) and $d\nu(\rho)/d\rho$ (\ref{98-1e}).\par
For the parameters in the above formula we adopt the values within the QCDSR approach \cite{NRSX} $\delta \simeq 1.9\ \rm{GeV}$, $\Lambda \simeq 0.65 - 1.0\ \rm{GeV}$, $\lambda \simeq - 0.2\ \rm{GeV}$, $<\overline{q} q>\ \simeq - \lambda^3$, $<\alpha_sGG >\ \simeq 0.12\ \rm{GeV}^4$, while for the function $\sigma(x)$ we consider the two limiting cases : $\sigma(w) = 1$ and $\sigma(w) = {1 \over 2} (w+1-\sqrt{w^2-1})$ (Figs. 8 and 9).\par

\subsubsection{Bounds on the derivatives}

For the case $\sigma(w) = 1$ we find that the lower bounds for the slope and the curvature (\ref{88e}) and (\ref{89e}) are satisfied, but the bound on the third derivative (\ref{90e}) is violated. For the case $\sigma(w) = {1 \over 2} (w+1-\sqrt{w^2-1})$ we find that the derivatives diverge at $w = 1$, and the lower bounds on the derivatives are trivially satisfied.

\subsubsection{Upper bound on the IW function}

We find that in general the QCDSR expression for the IW function (\ref{104-1e}) does not satisfy the upper bound (\ref{upper-6e}) $\xi(w) \leq  \left({2 \over 1+w}\right)^{3 \over 2}$. Although for the limiting case $\sigma(w) = {1 \over 2} (w+1-\sqrt{w^2-1})$ we find that it is satisfied, the bound is violated for the other limiting case $\sigma(w) = 1$.

\subsubsection{Positivity of the measure}

We see that the function $\eta_{QCDSR}(\tau)$ remains bounded, but not by 1, (Fig. 8), and we can compute its Fourier transform, that gives the measure $d\nu_{QCDSR}(\rho)/d\rho$ (Fig. 9).

\includegraphics[scale=1.]{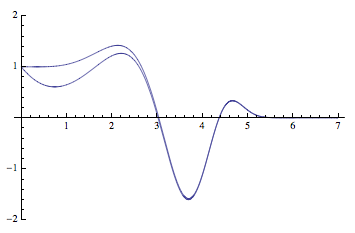}\\
Fig. 8. The function $\eta_{QCDSR}(\tau)$ (\ref{102-4e}) for the QCDSR formula (\ref{104-1e})(\ref{104-2e}) for the IW function in the cases $\sigma(x) = 1$ and $\sigma(x) = {1 \over 2} (x+1-\sqrt{x^2-1})$ (respectively upper and lower curves). 

\vskip 4 truemm

\includegraphics[scale=1.]{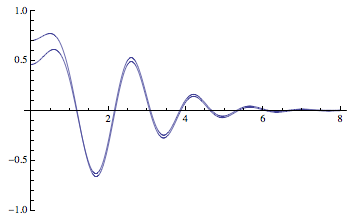}\\
Fig. 9. $d\nu_{QCDSR}(\rho)/d\rho$, Fourier transform of the function $\eta_{QCDSR}(\tau)$ in the cases $\sigma(x) = 1$ and $\sigma(x) = {1 \over 2} (x+1-\sqrt{x^2-1})$ (upper and lower curves at low $\rho$). 

\section{Bakamjian-Thomas relativistic quark model}

The Bakamjian-Thomas relativistic quark model \cite{BT,KP,T,CGNPSS} is a class of models with a fixed number of constituents in which the states are covariant under the Poincar\'e group. The model relies on an appropriate Lorentz boost of the eigenfunctions of a Hamiltonian describing the hadron spectrum at rest. From now on we use the abreviation BT for the Bakamjian-Thomas model, not be confused with the Buchm\"uller-Tye quarkonium potential model.\par 

We have proposed a formulation of this scheme for the meson ground states \cite{LOPR-1} and demonstrated the important feature that, in the heavy quark limit, the current matrix elements, when the current is coupled to the heavy quark, are {\it covariant}. We have extended this scheme to P-wave excited states \cite{MLOPR-1}.\par

Moreover, these matrix elements in the heavy quark limit exhibit Isgur-Wise (IW) scaling \cite{1r}. As demonstrated in \cite{LOPR-1,MLOPR-1}, given a Hamiltonian describing the spectrum, the model provides an unambiguous result for the Isgur-Wise functions, the elastic $\xi (w)$ \cite{1r} and the inelastic to P-wave states $\tau_{1/2}(w)$, $\tau_{3/2}(w)$ \cite{5r}.\par 

On the other hand, the sum rules (SR) in the heavy quark limit of QCD, like Bjorken \cite{4r,5r} and Uraltsev SR \cite{6r} are analytically satisfied in the model \cite{LOPR-2,MLOPR-2,LOPRM}, as well as SR involving higher derivatives of $\xi (w)$ at zero recoil \cite{LOR-1bis,LOR-2,LOR-3}.\par
  
In \cite{13bisr}, we have chosen the Godfrey-Isgur Hamitonian \cite{GI}, that gives a very complete description of the light $q\overline{q}$ and heavy $Q\overline{q}$ meson spectra in order to predict within the BT scheme the corresponding IW functions for the ground state and the excited states.\par

\subsection{Isgur-Wise function and positivity of the measure}

Let us now demonstrate that in the Bakamjian-Thomas relativistic quark model, the IW function implies a positive measure independently of the potential.\par
In this scheme, the IW function is given by the expression 
\beq
\label{105-1e}
\xi(v.v') = {1 \over 1+v.v'} \int {d{\vec p} \over p^0}\ {m(v.v'+1)+p.(v+v') \over \sqrt{(p.v+m)(p.v'+m)}}\ \varphi\left(\sqrt{(p.v')^2-m^2}\right)^* \varphi\left(\sqrt{(p.v)^2-m^2}\right) 
\eeq

\noindent with the wave function normalized according to
\beq
\label{105-2e}
\int {d{\vec p} \over p^0}\ |\varphi(|{\vec p}|)|^2 = 1 
\eeq

Let us first transform this expression in a convenient form (formula (\ref{105-13e}) below) that will allow us to compute the measure $d\nu(\rho)$ (\ref{98-2e}) of the decomposition of $\xi(w)$ in terms of irreducible IW functions  $\xi^\rho(w)$ (\ref{56-1e}) or (\ref{56-2e}).\par
Let us perform a change of integration variables :
\beq
\label{105-3e}
(p^1, p^2, p^3) \to (p^1, x = v.p, x' = v.p') 
\eeq

\noindent In this way, the arguments of $\varphi$ will not depend on $v$ and $v'$. Using the invariance of (\ref{105-1e}), we express $v, v'$ in terms of the variable $\tau$ (\ref{49e}) as follows :
\beq
\label{105-4e}
v = \left(\cosh(\tau/2), 0, 0, \sinh(\tau/2) \right) \qquad \qquad v' = \left(\cosh(\tau/2), 0, 0, - \sinh(\tau/2) \right)  
\eeq

\noindent one has $v.v' = \cosh(\tau)$ and
\beq
\label{105-5e}
x = \cosh(\tau/2) p^0 - \sinh(\tau/2) p^3\qquad \qquad x' = \cosh(\tau/2) p^0 + \sinh(\tau/2) p^3  
\eeq

\noindent The jacobian reads :
\beq
\label{105-6e}
{d{\vec p} \over p^0} = {1 \over \sinh(\tau)} {1 \over p^2}\ dp^1 dx dx'
\eeq

\noindent and (\ref{105-1e}) becomes (expression to be corrected below)
$$\xi(\cosh(\tau)) = {1 \over \cosh(\tau)+1} {1 \over \sinh(\tau)} \int {dp^1 \over |p^2|}\ dx dx'$$
\beq
\label{105-7e}
\times {m\left(\cosh(\tau)+1\right)+x+x' \over \sqrt{(x+m)(x'+m)}}\ \varphi\left(\sqrt{x'^2-m^2}\right)^*  \varphi\left(\sqrt{x^2-m^2}\right) 
\eeq

\noindent Using now (\ref{105-5e}) and $(p^2)^2 = (p^0)^2-(p^3)^2-(p^1)^2-m^2$ one gets the integration domain
\beq
\label{105-8e}
0 \leq (x'-e^{-\tau}x)(e^\tau x-x')-\sinh^2(\tau) m^2 \qquad \qquad  \qquad \
\eeq
\beq
\label{105-9e}
|p^1| \leq {\sqrt{(x'-e^{-\tau}x)(e^\tau x-x')-\sinh^2(\tau) m^2} \over \sinh(|\tau|)} \qquad \qquad \ \ 
\eeq
\beq
\label{105-10e}
p^2 = \pm {\sqrt{(x'-e^{-\tau}x)(e^\tau x-x')-\sinh^2(\tau) \left((p^1)^2-m^2\right)} \over \sinh(|\tau|)} 
\eeq

\noindent Let us first remark that (\ref{105-10e}) gives two values for $p^2$, and hence the integral (\ref{105-7e}) has to be multiplied by a factor 2 since both domains $p^2 \leq 0$ and $p^2 \geq 0$ correspond to the domain of $(p^1, x, x')$ given by (\ref{105-8e}) and (\ref{105-9e}).\par
On the other hand, (\ref{105-9e}) and (\ref{105-10e}) have the form $|p^1| \leq A$, $p^2 = \pm \sqrt{A^2-(p^1)^2}$, where $A$ can be read from (\ref{105-9e}) and hence one can compute the integral\par 
\noindent $\int {dp^1 \over |p^2|} = \int^A_{-A} {dp^1 \over \sqrt{A^2-(p^1)^2}} = \pi$. Using this value and multiplying (\ref{105-7e}) by the missing factor 2, we have 
$$\xi\left(\cosh(\tau)\right) = 2\pi {1 \over \cosh(\tau)+1} {1 \over \sinh(|\tau|)} \int \chi(0 \leq  (x'-e^{-\tau}x)(e^\tau x-x')-\sinh^2(\tau) m^2) dx dx'$$
\beq
\label{105-11e}
\times {m\left(\cosh(\tau)+1\right)+x+x' \over \sqrt{(x+m)(x'+m)}}\ \varphi\left(\sqrt{x'^2-m^2}\right)^*  \varphi\left(\sqrt{x^2-m^2}\right) 
\eeq

\noindent where the characteristic function $\chi(\mathcal{D})$ of a certain domain $\mathcal{D}$ is defined to be equal to 1 within the domain, and 0 outside.\par 

The equation (\ref{105-11e}) simplifies if we replace the variables of integration $x, x'$ by
\beq
\label{105-12e}
x = m \cosh(\alpha) \qquad \qquad \qquad x' = m \cosh(\alpha')
\eeq

\noindent since the constraint on $x, x'$ becomes $0 \leq (\cosh(\tau)-\cosh(\alpha'-\alpha))(\cosh(\alpha'+\alpha)-\cosh(\tau))$, or $|\alpha'-\alpha| \leq |\tau| \leq \alpha'+\alpha$ and (\ref{105-11e}) becomes
$$\xi\left(\cosh(\tau)\right) = 2\pi m^2 {1 \over \cosh(\tau)+1} {1 \over \sinh(|\tau|)} \int_0^\infty \int_0^\infty \chi(|\alpha'-\alpha| \leq |\tau| \leq \alpha'+\alpha) d\alpha d\alpha' $$
\beq
\label{105-13e}
\times \left(\cosh(\tau) + \cosh(\alpha) + \cosh(\alpha') + 1\right) f(\alpha')^*f(\alpha)
\eeq

\noindent where
\beq
\label{105-14e}
f(\alpha) = {\sinh(\alpha)\ \varphi(m \sinh(\alpha)) \over \sqrt{\cosh(\alpha)+1}}
\eeq

The normalization of the wave function $\varphi(\vec p)$ (\ref{105-2e}) translates into the condition for the function $f(\alpha)$ :
\beq
\label{105-15e}
4 \pi m^2 \int_0^\infty (\cosh(\alpha)+1)|f(\alpha)|^2 d\alpha = 1 
\eeq 

To compute the measure we need to go through formulas (\ref{91-1e})(\ref{93-2e})(\ref{98-2e}). We have first
$$\widehat{\xi}(\tau) = 2 \pi m^2 sgn(\tau) \int_0^\infty \int_0^\infty d\alpha d\alpha'\ \chi(|\alpha'-\alpha| \leq |\tau| \leq \alpha'+\alpha)$$
\beq
\label{105-16e}
\times \left(\cosh(\tau)+\cosh(\alpha)+\cosh(\alpha')+1\right) f(\alpha')^*f(\alpha) 
\eeq

\noindent and its derivative is given by
$${d \over d\tau}\ \widehat{\xi}(\tau) = 2 \pi m^2 \int_0^\infty \int_0^\infty d\alpha d\alpha'\ f(\alpha')^*f(\alpha)$$
$$\times\ ((\delta(|\alpha'-\alpha|-|\tau|)-\delta(\alpha'+\alpha-|\tau|))(\cosh(\tau)+\cosh(\alpha)+\cosh(\alpha')+1)$$
\beq
\label{105-17e}
+ \sinh(|\tau|)\ \chi(|\alpha'-\alpha| \leq |\tau| \leq \alpha'+\alpha))) 
\eeq

\noindent This expression simplifies to
$${d \over d\tau}\ \widehat{\xi}(\tau) = 2 \pi m^2 \int_0^\infty \int_0^\infty d\alpha d\alpha'\ f(\alpha')^*f(\alpha)$$
$$\times\ (4 \cosh(\tau/2)\ (\delta(|\alpha'-\alpha|-|\tau|)-\delta(\alpha'+\alpha-|\tau|)) \cosh(\alpha'/2) \cosh(\alpha/2)$$
\beq
\label{105-18e}
+ \sinh(|\tau|)\ \chi(|\alpha'-\alpha| \leq |\tau| \leq \alpha'+\alpha)))
\eeq

\noindent and finally one gets the function
$$\eta(\tau) = 2 \pi m^2 \int_0^\infty \int_0^\infty d\alpha d\alpha'\ f(\alpha')^*f(\alpha)$$
$$\times (2 (\delta(|\alpha'-\alpha|-|\tau|)-\delta(\alpha'+\alpha-|\tau|)) \cosh(\alpha'/2) \cosh(\alpha/2)$$
\beq
\label{105-19e}
+ \sinh(|\tau|/2)\ \chi(|\alpha'-\alpha| \leq |\tau| \leq \alpha'+\alpha))
\eeq

We have now to compute the Fourier transform (\ref{98-2e}) of this function. Let us consider the first term of (\ref{105-19e}) :
\beq
\label{105-20e}
\int_{-\infty}^{+\infty} e^{i\rho \tau} (\delta(|\alpha'-\alpha|-|\tau|)-\delta(\alpha'+\alpha-|\tau|)) d\tau = -4 \sinh(i\rho\alpha)\sinh(i\rho\alpha')
\eeq

\noindent and the second term :
\beq
\label{105-21e}
\int_{-\infty}^{+\infty} e^{i\rho \tau} \sinh(|\tau|/2)\ \chi(|\alpha'-\alpha| \leq |\tau| \leq \alpha'+\alpha) =
\eeq
$$2\left({1 \over i\rho+{1 \over 2}} \sinh((i\rho+{1 \over 2})\alpha') \sinh((i\rho+{1 \over 2})\alpha) - {1 \over i\rho-{1 \over 2}} \sinh((i\rho-{1 \over 2})\alpha') \sinh((i\rho-{1 \over 2})\alpha)\right)$$

\noindent and we finally obtain the following expression for the measure
$${d\nu(\rho) \over d\rho} = 2 m^2 \int_0^\infty \int_0^\infty d\alpha d\alpha'\ f(\alpha')^*f(\alpha) (-4 \sinh(i\rho\alpha')\cosh(\alpha'/2) \sinh(i\rho\alpha)\cosh(\alpha/2)$$
\beq
\label{105-22e}
+ {1 \over i\rho+{1 \over 2}} \sinh((i\rho+{1 \over 2})\alpha') \sinh((i\rho+{1 \over 2})\alpha) - {1 \over i\rho-{1 \over 2}} \sinh((i\rho-{1 \over 2})\alpha') \sinh((i\rho-{1 \over 2})\alpha))
\eeq 

What needs to be demonstrated now is that indeed this measure is positive, ${d\nu(\rho) \over d\rho} \geq 0$.
To this purpose, let us define two functions, transformed of $f(\alpha)$ : 
\beq
\label{105-23e}
g_\pm(\rho) = \int_0^{\infty} \sinh ((i\rho \pm {1 \over 2})\alpha) f(\alpha) d\alpha
\eeq

\noindent in terms of which (\ref{105-22e}) becomes
\beq
\label{105-23bise}
{d\nu(\rho) \over d\rho} = 2m^2 \left( |g_+(\rho)+g_-(\rho)|^2 - {g_-(\rho)^*g_+(\rho) \over {i\rho + {1 \over 2}}} + {g_-(\rho)g_+(\rho)^* \over {i\rho - {1 \over 2}}} \right)
\eeq

\noindent and the measure can be expressed as a modulus squared
\beq
\label{105-24e}
{d\nu(\rho) \over d\rho} = |h(\rho)|^2
\eeq 

\noindent where the function $h(\rho)$ is given by the expression :
\beq
\label{105-25e}
h(\rho) = - \sqrt{2} m\ {1 \over \sqrt{\rho^2 + {1 \over 4}}} \left((i\rho - {1 \over 2}) g_+(\rho) + (i\rho + {1 \over 2}) g_-(\rho) \right) 
\eeq 

We conclude that the measure $d\nu(\rho)/d\rho$ is positive.\par
Moreover, one must notice that ${d\nu(\rho) \over d\rho}$ is a function, and therefore it does not contain discrete $\delta$-function terms. This follows from the fact that, according to (\ref{105-23e}), $g_\pm(\rho)$ are Fourier transforms of functions that, from (\ref{105-15e}), are square integrable and therefore are themselves functions (square integrable).\par
So, not all possible IW functions $\xi(w)$ are obtained in the BT models. For instance, the so-called BPS limit for the slope $-\xi'(1) = {3 \over 4}$, 
leading to the function (\ref{57-4e}) cannot be obtained.

\subsection{Lorentz group representation for the BT model}

We will begin with a short description of what was exposed in detail for the baryon case $j = 0$ \cite{LOR-1}.\par
The starting point is an arbitrary unitary representation $U$ of the Lorentz group $SL(2,C)$ in an arbitrary Hilbert space $\mathcal{H}$. To have the meson states and define the Isgur-Wise functions it is moreover necessary that $\mathcal{H}$ is provided with a mass operator $M$ that commutes with the rotations, i.e. with the subgroup $SU(2)$ of $SL(2,C)$. The eigenvalues and eigenvectors of $M$ will give the spectrum and eigenfunctions of the mesons at rest.\par
The Hilbert space $\mathcal{H}$ will describe the states of the light cloud and $M$ will describe the effect of the heavy quark at rest on the latter. Hence, the states of the light cloud that correspond to the hadrons (for the heavy quark at rest) are the eigenstates of $M$.\par
The first step is to determine the irreducible representations of spin $j$ of the restriction of $U$ to $SU(2)$, with their standard bases $|j, \mu >$.\par
When one has the states of the light cloud of a hadron at rest $v_0 = (1, \vec{0})$, the states at arbitrary velocity $v$ are obtained from $U(\Lambda)$, with Lorentz transformation $\Lambda$ transforming $v_0$ into $v$. However, we need more specifically the states $|j, v, \epsilon >$ where the spin is specified by a polarization tensor $\epsilon$, that transform in a covariant way as follows :
\beq
\label{105-26e}
U(\Lambda)|j, v, \epsilon >\ = |j, \Lambda v, \Lambda \epsilon >
\eeq

\noindent These states are given by the following formula :
 \beq
\label{105-27e}
|j, v, \epsilon >\ = \sum_\mu < \epsilon^\mu | B^{-1}_v \epsilon > U(B_v) |j, \mu >
\eeq

Let us precise that the tensors $\epsilon$ at velocity $v$ constitute a vector space $\mathcal{E}_{j,v}$ of dimension $2j + 1$, and $\Lambda \in SL(2,C)$ applies $\mathcal{E}_{j,v}$ on $\mathcal{E}_{j,\Lambda v}$ and that on $\mathcal{E}_{j,v_0}$ acts the representation $j$ of $SU(2)$. Then in (\ref{105-27e}) $(\epsilon^\mu)_{-j \leq \mu \leq j}$ is a standard basis of $\mathcal{E}_{j,v_0}$, one has $B_v^{-1} \epsilon \in \mathcal{E}_{j,v_0}$ and $< \epsilon^\mu | B^{-1}_v \epsilon >\ = (B_v^{-1}\epsilon)_\mu$ are the components of $B_v^{-1}\epsilon$ on this basis. On the other hand, $B_v \in SL(2,C)$ is the boost $v_0 \to v$, but (\ref{105-27e}) gives the same state $|j, v, \epsilon >$ if $B_v$ is replaced  by any $\Lambda : v_0 \to v$.\par
The second step is therefore to compute the states defined by (\ref{105-27e}). Finally, what remains is to compute the scalar products $< j', v', \epsilon' |j, v, \epsilon >$. Because of (\ref{105-26e}) and the unitarity of $U$, these scalar products satisfy $< j', v', \epsilon' |j, v, \epsilon >\ =\ < j', \Lambda v', \Lambda \epsilon' |j, \Lambda v, \Lambda \epsilon >$, i.e. are functions of $v, \epsilon, v', \epsilon'$, invariant under Lorentz transformations. The Isgur-Wise functions are then the coefficients, functions of only $v.v'$, in the expansion of these scalar products on a basis of these invariants.\par
We will now apply this program to a particular representation of $SL(2,C)$ and obtain in this way the IW functions in the BT model, that were computed elsewhere. We do not need to specify the mass operator $M$.\par

\subsubsection{Description of the Lorentz group representation}

The representation of $SL(2,C)$ that we consider is the one obtained from a spin $1/2$ particle by restriction of the Poincar\' e group to the Lorentz group. The Hilbert space $\mathcal{H}$ is $L^2_{C^2}(H_m,d\mu(p))$ of the functions on the mass hyperboloid $H_m = \{p \in R^4\ |\ p^2 = m^2,\ p^0 > 0\}$, with values in the space $C^2$ of the unitary representation $D^{1/2}$ of $SU(2)$ of spin $1/2$, with the scalar product : 
\beq
\label{105-28e}
<\psi' |\psi >\ = \int d\mu(p) <\psi'(p) |\psi(p) >
\eeq

\noindent where $d\mu(p)$ is the invariant measure on the mass hyperboloid
\beq
\label{105-29e}
d\mu(p) = {d^3\vec p \over p^0}
\eeq

\noindent and the action of $\Lambda \in SL(2,C)$ in $\mathcal{H}$ is given by
\beq
\label{105-29bise}
(U(\Lambda) \psi)(p) = D^{1/2}({\bf R}(\Lambda, p)) \psi(\Lambda^{-1} p)
\eeq

\noindent where the Wigner rotation ${\bf R}(\Lambda, p) \in SU(2)$ is
\beq
\label{105-30e}
{\bf R}(\Lambda, p) = B_p^{-1} \Lambda B_{\Lambda^{-1}p}
\eeq

\noindent where $B_p \in SL(2,C)$ is the boost $(m, \vec{0}) \to p$.\par
The check of the group law $U(\Lambda') U(\Lambda) = U(\Lambda' \Lambda)$ follows from a simple calculation, and unitarity comes from the unitarity of $D^{1/2}$ and the invariance of the measure $d\mu(p)$.

\subsubsection{States $j$ of the light cloud for the heavy quark at rest}

We do not have to specify here the mass operator $M$ (for example it can be the hamiltonian of Godfrey-Isgur \cite{GI} in the heavy quark limit). We need simply to describe the irreducible representations of spin $j$ of the restriction to $SU(2)$, with their standard bases.\par
For a rotation $\Lambda = R \in SU(2)$, the transformation (\ref{105-29e}) reduces to
\beq
\label{105-31e}
(U(R) \psi)(p) = D^{1/2}(R) \psi(R^{-1} p)
\eeq

\noindent because the Wigner rotation is simply $R$
\beq
\label{105-32e}
{\bf R}(R,p) = R
\eeq

\noindent This can be seen using the following characterization of the boost :
\beq
\label{105-33e}
\Lambda (m, \vec{0}) = p, \qquad \Lambda = \Lambda^\dagger, \qquad \Lambda > 0 \qquad \Leftrightarrow \qquad \Lambda = B_p  
\eeq

\noindent that implies
\beq
\label{105-34e}
R B_{R^{-1}p} R^{-1} = B_p
\eeq

\noindent Therefore, from (\ref{105-31e}), the calculation is reduced to the combination of an orbital angular momentum $L$ with a spin ${1 \over 2}$ described by a Pauli spinor $\chi$.\par
For each value of $j$ one has two families of solutions $(L = j \pm {1 \over 2})$ of opposite parity $(-1)^L$ :
\beq
\label{105-35e}
\varphi^{(L,j,\mu)}(p) = \sqrt{4 \pi}\ (Y_L \chi)_j^\mu({\hat p})\ \varphi^{(L,j)}(|\vec p|)\ ,
\eeq
$$(Y_L \chi)_j^\mu({\hat p}) = \sum_{M, \mu'} < j, \mu | L, M, {1 \over 2}, \mu' > Y_L^M({\hat p})\ \chi^{\mu'}$$ 

\noindent that depend on the radial function $\varphi^{(L,j})(|\vec p|)$ normalized by
 \beq
\label{105-36e}
\int {d^3{\vec p} \over p^0}\ |\varphi^{(L,j)}(|\vec p|)|^2 = 1
\eeq

Following formula (\ref{105-27e}), the next step is the calculation of the wave functions of the light cloud $\varphi^{(L,j,v,\epsilon)}(p)$ for a velocity $v$ and a polarization tensor $\epsilon$, starting from the functions $\varphi^{(L,j,\mu)}(p)$ given by (\ref{105-35e}). This is enormously simplified if one uses a representation of $SL(2,C)$ {\it equivalent} to the precedent one, expressed in terms of spinors and Dirac matrices.

\subsubsection{Representation in terms of Dirac spinors and matrices}

Let us introduce now the space $\mathcal{H}'$, another way of describing the space $\mathcal{H}$, constituted of functions on the hyperboloid $H_m$, taking values at the point $p \in H_m$ in the sub-space of $C^4$ constituted by the Dirac spinors which satisfy
\beq
\label{105-36bise}
({/\hskip - 2 truemm p}-m) u = 0
\eeq

\noindent since in the BT model the light quark is on-shell. The scalar product is
\beq
\label{105-37e}
<\psi' |\psi >\ = \int d\mu(p)\ {\bar \psi}'(p) \psi(p) \qquad \qquad ({\bar \psi}(p) = \psi^\dagger (p) \gamma^0)
\eeq

\noindent and the action of $\Lambda \in SL(2,C)$ is given by :
\beq
\label{105-38e}
(U(\Lambda) \psi)(p) = D(\Lambda) \psi(\Lambda^{-1} p)
\eeq

\noindent where $D(\Lambda)$ is the Dirac matrix of the Lorentz transformation $\Lambda$ :
\beq
\label{105-39e}
D(\Lambda) = {1 \over 2} 
\left( \begin{array}{cc}
\Lambda+\Lambda^{\dagger{-1}} & \Lambda-\Lambda^{\dagger{-1}} \\
\Lambda-\Lambda^{\dagger{-1}} & \Lambda+\Lambda^{\dagger{-1}}
\end{array} \right)
\eeq

The unitary tranformation $V$ : $\mathcal{H} \to \mathcal{H}'$ 
\beq
\label{105-40e}
\psi(p) = (V\varphi)(p)
\eeq

\noindent that implements the equivalence is given by
\beq
\label{105-41e}
(V \varphi)(p) = D(B_p) Q^\dagger \varphi(p)
\eeq
\beq
\label{105-42e}
(V^{-1} \psi)(p) = Q D(B_p^{-1}) \psi(p)
\eeq 

\noindent where the operators $Q$ and $Q^\dagger$ make the connection between the four-component spinors and the two-component ones :
\beq
\label{105-43e}
Q \left( \begin{array}{c} \chi_1 \\ \chi_2 \end{array} \right) = \chi_1 \qquad \qquad Q^\dagger \chi = \left( \begin{array}{c} \chi \\ 0 \end{array} \right) 
\eeq

\noindent Let us collect some identities used to establish the equivalence :
$$(a) \ \ \ QQ^\dagger = 1\ ; \qquad \qquad \qquad \ \ \ \ \ $$ 
$$(b) \ \ \ Q^\dagger Q = {1+\gamma^0 \over 2}\ ; \qquad \qquad \ \ \ \ $$
$$(c) \ \ \ (\gamma^0-1) Q^\dagger = 0\ ; \qquad \qquad \ \ \ $$
\beq
\label{105-43bise}
\qquad \qquad \qquad (d) \ \ \  Q^\dagger D^{1/2}(R) Q = Q^\dagger Q D(R) \ \ (R \in SU(2))\ ; \
\eeq
$$(e) \ \ \ D(\Lambda') D(\Lambda) = D(\Lambda' \Lambda)\ ; \qquad$$
$$ \qquad \ \ \ (f) \ \ \ D(\Lambda) {/\hskip - 2 truemm a} D(\Lambda^{-1}) =\ {/\hskip - 3.5 truemm \Lambda a}\ ; \qquad \qquad \qquad$$
$$(g) \ \ \ D(\Lambda)^\dagger = \gamma^0 D(\Lambda ^{-1}) \gamma^0\ . \qquad$$

Applying $V^{-1}$ to $\psi = V\phi$ one finds, using relation (d) and then (a), $(V^{-1} \psi)(p) = \phi(p)$ and therefore $V^{-1} V = 1$. Next, if  $\psi = V \phi$ one has $({/\hskip - 2 truemm p} -m)\psi(p) = ({/\hskip - 2 truemm p} -m)D(B_p)Q^\dagger\phi(p) = 0$, using (d) and (f) and then (c).\par
Applying $V$ to $\phi = V^{-1} \psi$, with $\psi(p)$ satisfying $({/\hskip - 2 truemm p} -m)\psi(p) = 0$, one obtains $(V \phi)(p) = D(B_p) Q^\dagger Q D(B_p^{-1})\psi(p) = \psi(p)$, using (b) then (f) and finally $({/\hskip - 2 truemm p} -m)\psi(p) = 0$, one gets also $VV^{-1} = 1$.\par 

To establish the unitarity of $V$ one needs to show that $< V^{-1}\psi' | V^{-1}\psi >\ =\ < \psi' | \psi >$, where on the left one has the scalar product (\ref{105-28e}), and on the right the scalar product (\ref{105-37e}). One has :
$$< V^{-1}\psi' | V^{-1}\psi >\ = \int d\mu(p) \psi'(p)^\dagger D(B_p^{-1})^\dagger P^\dagger P D(B_p^{-1}) \psi(p)$$
\beq
\label{105-47e}
= \int d\mu(p) \overline{\psi'}(p) \psi(p) =\ < \psi' | \psi > \qquad \qquad \qquad \qquad \qquad  
\eeq

\noindent using identities (b) and then (g), (f) and finally $({/\hskip - 2 truemm p} -m)\psi(p) = 0$. This establishes also that the scalar product (\ref{105-37e}) is indeed positive definite. \par
Finally, it remains to verify that the transformation law $V U(\Lambda) V^{-1}$ in $\mathcal{H}'$, transported from $U(\Lambda)$ in $\mathcal{H}$, given by (\ref{105-29bise}) and (\ref{105-30e}), by $V$ is given by (\ref{105-38e}) :
$$(VU(\Lambda)V^{-1} \psi)(p) = D(B_p) Q^\dagger D^{1/2}({\bf R} (\Lambda, p)) Q D(B^{-1}_{\Lambda^{-1} p}) \psi(\Lambda^{-1} p)$$
\beq
\label{105-48e}
= D(\Lambda)\ \ { {/\hskip - 5 truemm \Lambda^{-1} p} + m \over 2m}\ \psi(\Lambda^{-1} p) = D(\Lambda) \psi(\Lambda^{-1} p)
\eeq

\noindent using (d), then (\ref{105-30e}) and (e), then (b) and (f), then $({/\hskip - 2 truemm p} -m)\psi(p) = 0$.

\subsubsection{States $j$ of the light cloud in the Dirac representation}

Concerning the states $| j, \mu >$ in $\mathcal{H}'$, they are obtained from the states $| j, \mu >$ in $\mathcal{H}$ given by (\ref{105-35e}) by applying the transformation $V$ given by (\ref{105-41e}), i.e. $\psi^{(L,j,\mu)}(p) = V \varphi^{(L,j,\mu)}(p)$, that gives
\beq
\label{105-49e}
\psi^{(L,j,\mu)}(p) = \sqrt{4 \pi}\ D(B_p) \left( \begin{array}{c} (Y_L \chi)_j^\mu (\hat p) \\ 0 \end{array} \right)  \varphi^{(L,j)}(|\vec p|)
\eeq

\noindent or
\beq
\label{105-50e}
\psi^{(L,j,\mu)}(p) = \sqrt{4 \pi} {{/\hskip - 2 truemm p} + m \over \sqrt{2m(p^0+m)}} \left( \begin{array}{c} (Y_L \chi)_j^\mu (\hat p) \\ 0 \end{array} \right) \varphi^{(L,j)}(|\vec p|)
\eeq

\noindent where we have used 
\beq
\label{105-51e}
D(B_p) = { m + {/\hskip - 2 truemm p} \gamma^0 \over \sqrt{2m(p^0+m)}}
\eeq

We will see that the calculation of (\ref{105-27e}) is simple when the $\mu$ dependence of $| j, \mu >$ appears under the form $(Y_{j-1/2} \chi)_j^\mu$, as it is the case with (\ref{105-50e}) for $L = j - 1/2$. In the case $L = j + 1/2$ one can also express $\psi^{(L,j,\mu)}$ in terms of $(Y_{j-1/2} \chi)_j^\mu$ by using the identity :
\beq
\label{105-52e}
(Y_{j+1/2} \chi)_j^\mu (\hat p) = - ({\vec \sigma}.\hat p) (Y_{j-1/2} \chi)_j^\mu (\hat p)
\eeq

\noindent and from (\ref{105-52e}) one gets, after some algebra :
\beq
\label{105-53e}
({/\hskip - 2 truemm p} + m) \left( \begin{array}{c} (Y_{j+1/2} \chi)_j^\mu (\hat p) \\ 0 \end{array} \right) = - \sqrt{{p^0+m \over p^0-m}}\ \gamma_5 ({/\hskip - 2 truemm p} - m) \left( \begin{array}{c} (Y_{j-1/2} \chi)_j^\mu (\hat p) \\ 0 \end{array} \right) 
\eeq

\noindent and we have, finally 
\beq
\label{105-54e}
\psi^{(j-1/2,j,\mu)}(p) = \sqrt{4 \pi} {{/\hskip - 2 truemm p} + m \over \sqrt{2m(p^0+m)}} \left( \begin{array}{c} (Y_{j-1/2} \chi)_j^\mu (\hat p) \\ 0 \end{array} \right) \varphi^{({j-1/2},j)}(|\vec p|)
\eeq
\beq
\label{105-55e}
\psi^{(j+1/2,j,\mu)}(p) = - \sqrt{4 \pi} \gamma_5 {{/\hskip - 2 truemm p} - m \over \sqrt{2m(p^0-m)}} \left( \begin{array}{c} (Y_{j-1/2} \chi)_j^\mu (\hat p) \\ 0 \end{array} \right) \varphi^{({j+1/2},j)}(|\vec p|)
\eeq

\subsubsection{States for arbitrary velocity and polarization tensor}

We can now go to the second step, the calculation using (\ref{105-27e}) of the wave functions $\psi^{(j \pm 1/2,j,v,\epsilon)}(p)$ of the states $| j, v, \epsilon >$ using the wave functions $\psi^{(j \pm 1/2,j,\mu)}(p)$ of the states $| j, \mu >$, given by (\ref{105-54e}) and (\ref{105-55e}), with $U(\Lambda)$ given by (\ref{105-38e}). We have then to compute
\beq
\label{105-56e}
\psi^{(j \pm 1/2,j,v,\epsilon)}(p) = \sum_\mu < \epsilon^\mu | B^{-1}_v \epsilon > D(B_v) \psi^{(j \pm 1/2,j,\mu)}(B_v^{-1} p) 
\eeq 

\noindent To do that, we need some precisions on the polarization tensors.\par 
For $j$ half-integer, they constitute the subspace $\mathcal{E}_{j,v}$ (dependent on the velocity $v$) of $(C^4)^{\otimes (j-1/2)} \otimes C^4$ of the tensors $\epsilon_\alpha ^{\mu_1,...,\mu_{j-1/2}}$ that satisfy the conditions :
$$(a)  \ \ \ \rm{symmetry\ under\ permutation\ of\ the}\ \mu\ \rm{indices}\ ; $$ 
$$(b)  \ \ \ \rm{null\ trace,\ i.e.}\ g_{\mu_1,\mu_2} \epsilon_\alpha ^{\mu_1,...,\mu_{J-1/2}} = 0\ (j \geq {5 \over 2})\ ; \ \ \ $$
\beq
\label{105-56bise} 
(c)  \ \ \ (\gamma_{\mu_1})_{\alpha,\beta} \epsilon_\beta ^{\mu_1,...,\mu_{J-1/2}}\ (j \geq {3 \over 2})\ ; \qquad \qquad \qquad \qquad
\eeq
$$(d)  \ \ \ v_{\mu_1} \epsilon_\alpha ^{\mu_1,...,\mu_{J-1/2}}\ (j \geq {3 \over 2})\ ; \qquad \qquad \qquad \qquad \ \ \ $$ 
$$(e)  \ \ \ ({/\hskip - 2 truemm v}-1)_{\alpha,\beta} \epsilon_\beta ^{\mu_1,...,\mu_{J-1/2}} = 0\ . \qquad \qquad \qquad \qquad \ \ $$ 

The Lorentz transformation of the polarization tensor is the following :
\beq
\label{105-57e}
(\Lambda \epsilon)_\alpha ^{\mu_1,...,\mu_{j-1/2}} = \Lambda_{\nu_1}^{\mu_1} ...\ \Lambda_{\nu_{j-1/2}}^{\mu_{j-1/2}} D(\Lambda)_{\alpha,\beta} \epsilon_\beta ^{\nu_1,...,\nu_{j-1/2}} 
\eeq

\noindent One sees that $\Lambda$ transforms $\mathcal{E}_{j,v}$ into $\mathcal{E}_{j,\Lambda v}$, that $\mathcal{E}_{j,v}$ is obtained from the space at rest $\mathcal{E}_{j,v_0}$ by $\Lambda$ when $\Lambda v_0 = v$, and that $\mathcal{E}_{j,v_0}$ applies to itself by rotations. Noting that in (\ref{105-56e}) the tensors $\epsilon^\mu$ and $B_v^{-1} \epsilon$ are in $\mathcal{E}_{j,v_0}$, it is clear that the sum in (\ref{105-56e}) requires to consider the polarisation tensors at zero velocity $v_0 = (1, {\vec 0})$.\par
For the tensors at zero velocity, the condition (d) means that any component with some $\mu = 0$ vanishes, and condition (e) means that any component where the index $\alpha$ is equal to 3 or 4 vanishes. Thus, keeping the other components, $\mathcal{E}_{j,v_0}$ identifies with $(C^3)^{\otimes (j-1/2)} \otimes C^2$ that, from the point of view of rotations, is the tensor product of $j-1/2$ angular momenta equal to 1 and one angular momentum $1/2$. Then, conditions (a), (b) and (c) mean simply that this subspace is the one where these angular momenta add to the maximal possible value $j$.

For the polarization tensors one has, at rest, the following identity :
\beq
\label{105-58e}
\sum_\mu < \epsilon^\mu | \epsilon > (Y_{j-1/2} \chi)_j^\mu ({\hat p}') = N_{j-1/2} {1 \over \sqrt{4 \pi}} \sum_{i_1...i_{j-1/2}} {\hat p}'^{i_1} ... {\hat p}'^{i_{j-1/2}}\ \epsilon ^{i_1 ... i_{j-1/2}}
\eeq

\noindent with
\beq
\label{105-59e}
N_L = {\sqrt{(2L+1)!} \over 2^{L/2} L!}
\eeq

To demonstrate these formulas one has first to establish the relation
\beq
\label{105-59-1e}
Y_L^M({\hat p}) = {N_L \over \sqrt{4 \pi}} \sum_{i_1...i_L} {\hat p}^{i_1} ... {\hat p}^{i_L} \left(\epsilon^M \right)^{i_1 ... i_L}
\eeq

\noindent where $\epsilon^M$ form a standard basis of polarization tensors (tridimensional at zero velocity) for an integer spin $L$. These $\epsilon^M$ are obtained by coupling  $L$ spins equal to 1 to the maximum value $L$.\par 
The Clebsch-Gordan coefficients  that couple two spins $J$ and $J'$ to the maximum value $J+J'$ are given by
\beq
\label{105-59-2e}
< J, J', M, M' | J+J', M+M' >\ = {C(J, M)C(J', M') \over C(J+J', M+M')} 
\eeq

\noindent with
\beq
\label{105-59-3e}
C(J, M) = \sqrt{{(2J)! \over (J-M)!(J+M)!}}
\eeq

\noindent Then one gets
\beq
\label{105-59-4e}
\epsilon^M = \sum_{m_1+ ... m_L = M} {C(1,m_1) ... C(1,m_L) \over C(L,M)}\ e^{m_1} ... e^{m_L}
\eeq

\noindent where the $e^m$ form a standard basis
\beq
\label{105-59-5e}
\epsilon^{+1} = - {e^1+i e^2 \over \sqrt{2}}, \qquad e^0 = e^3, \qquad \epsilon^{-1} = {e^1-i e^2 \over \sqrt{2}}
\eeq

\noindent Let us now consider the generating function of the $Y_L^M$ :
$$\sum_M {L! \over \sqrt{(L-M)!(L+M)!}} Y_L^M(\hat{p}) s^{L+M} $$
\beq
\label{105-59-6e}
= \sqrt{{2L+1 \over 4 \pi}} \left({\hat{p}^1-i\hat{p}^2 \over 2} + \hat{p}^3 s - {\hat{p}^1+i\hat{p}^2 \over 2} s^2 \right)^L 
\eeq

\noindent and let us compute the generating function of the r.h.s. of (\ref{105-59-1e}). Using (\ref{105-59-4e}) one finds 
$$\sum_M {L! \over \sqrt{(L-M)!(L+M)!}} {N_L \over \sqrt{4 \pi}} \sum_{i_1 ... i_L} {\hat p}^{i_1} ... {\hat p}^{i_L} \left(\epsilon^M \right)^{i_1 ... i_L} s^{L+M}$$
\beq
\label{105-59-7e}
= {N_L \over \sqrt{4 \pi}} {L! \over \sqrt{(2L)!}} \left( \sum_m C(1,m) (\hat{p}.e^m) s^{1+m} \right)^L
\eeq

\noindent and, taking into account (\ref{105-59-5e}), one has
$$\sum_m C(1,m) (\hat{p}.e^m) s^{1+m} = (\hat{p}.e^{-1}) + \sqrt{2} (\hat{p}.e^0)s + (\hat{p}.e^{+1}) s^2$$
\beq
\label{105-59-8e}
= \sqrt{2} \left({\hat{p}^1-i\hat{p}^2 \over 2} + \hat{p}^3 s - {\hat{p}^1+i\hat{p}^2 \over 2} s^2 \right)
\eeq

\noindent and one sees that both generating functions are identical provided $N_L$ is given by (\ref{105-59e}). This establishes the relation (\ref{105-59-1e}) with (\ref{105-59e}), and from it one easily obtains (\ref{105-58e}). This ends the demonstration of formulas (\ref{105-58e}) and (\ref{105-59e}).

Taking $p' = B_v^{-1} p$, the identity (\ref{105-58e}) allows easily to make the sum over $\mu$ in (\ref{105-56e}) for $\psi^{(j \pm 1/2,j,\mu)}(p)$ given by (\ref{105-54e}) and (\ref{105-55e}). Indeed, (\ref{105-58e}) gives :
$$\sum_\mu < \epsilon^\mu | B_v^{-1} \epsilon > (Y_{j-1/2} \chi)_j^\mu ({\hat p}')$$
\beq
\label{105-60e}
= N_{j-1/2} {1 \over \sqrt{4 \pi}} \sum_{i_1,...i_{j-1/2}}\hat{p}^{i_1}...\ \hat{p}^{i_{j-1/2}}\ (B_v^{-1} \epsilon) ^{i_1 ... i_{j-1/2}}
\eeq

\noindent and using $D(\Lambda') D(\Lambda) = D(\Lambda' \Lambda)$ and
\beq
\label{105-61e}
(B_v^{-1} p)^0 = p.v\,  \qquad \qquad |{\vec {B_v^{-1} p}}| = \sqrt{(p.v)^2-m^2}  
\eeq

\noindent one gets for $\psi^{(L,j,v,\epsilon)}(p)$, omitting the spinorial index :
$$\psi^{(j - 1/2,j,v,\epsilon)}(p) = (-1)^{j-1/2} {N_{j-1/2} \over \sqrt{2m(p.v+m)}}$$
\beq
\label{105-62e}
({/\hskip - 2 truemm p}+m)\ p_{\mu_1} ... p_{\mu_{j-1/2}} \epsilon^{\mu_1 ... \mu_{j-1/2}}\ {\varphi^{(j-1/2,j)}(\sqrt{(p.v)^2-m^2}) \over (\sqrt{(p.v)^2-m^2})^{j-1/2}}
\eeq

\noindent and
$$\psi^{(j + 1/2,j,v,\epsilon)}(p) = -\ (-1)^{j-1/2} {N_{j-1/2} \over \sqrt{2m(p.v-m)}}\ \gamma_5$$
\beq
\label{105-63e}
({/\hskip - 2 truemm p}-m)\ p_{\mu_1} ... p_{\mu_{j-1/2}}\epsilon^{\mu_1 ... \mu_{j-1/2}}\ {\varphi^{(j+1/2,j)}(\sqrt{(p.v)^2-m^2}) \over (\sqrt{(p.v)^2-m^2})^{j-1/2}}
\eeq

\subsubsection{Isgur-Wise functions}

We will now consider three cases of physical interest for which, in the scalar product of states, a single IW function is involved, namely the ground state elastic case $\{ j = 1/2, L = 0 \to j = 1/2, L = 0 \}$ and the ground state to $L = 1$ states $j = 1/2, 3/2$ : $\{ j = 1/2, L = 0 \to j = 1/2, L = 1$ \},  \{$j = 1/2, L = 0 \to j = 3/2, L = 1 \}$.\par

\vskip 5 truemm
\underline{Elastic case $ j = 1/2, L = 0 \to j = 1/2, L = 0$ } \par
\vskip 5 truemm

For the ground state IW function $\xi(w)$ one must compute the overlap (for $j = 1/2$ the tensor $\epsilon$ is just a spinor) 
\beq
\label{105-64e}
<\psi^{(0,1/2,v',\epsilon ')} |\psi^{(0,1/2,v,\epsilon)} >\ = \xi(w)\ \overline{\epsilon '}\epsilon
\eeq

\noindent where 
\beq
\label{105-65e}
\psi^{(0,1/2,v,\epsilon)}(p) = {1 \over \sqrt{2m(p.v+m)}} ({/\hskip - 2 truemm p}+m)\ \epsilon\ \phi^{(0,1/2)}(\sqrt{(p.v)^2-m^2)}
\eeq

\noindent With the scalar product defined by (\ref{105-28e}) with the measure (\ref{105-29e}) one obtains
$$<\psi^{(0,1/2,v',\epsilon ')} |\psi^{(0,1/2,v,\epsilon)} >\ = \int {d^3\vec p \over p^0} {1 \over \sqrt{p.v+m}} {1 \over \sqrt{p.v'+m}}$$
\beq
\times\ \overline{\epsilon '}({/\hskip - 2 truemm p}+m)\epsilon\ \varphi^{(0,1/2)}(\sqrt{(p.v')^2-m^2})^* \varphi^{(0,1/2)}(\sqrt{(p.v)^2-m^2}) 
\label{105-66e}
\eeq

\noindent parametrizing the integrals of (\ref{105-66e}) under the form
\beq
A(w) = \int {d^3\vec p \over p^0}\ F(p,v,v') \qquad \qquad \qquad \qquad 
\label{105-67e}
\eeq
\beq
B(w)v^\mu + C(w) v'^\mu = \int {d^3\vec p \over p^0}\ F(p,v,v')\ p^\mu \ 
\label{105-68e}
\eeq

\noindent where
\beq
F(p,v,v') = {\varphi^{(0,1/2)}(\sqrt{(p.v')^2-m^2})^* \varphi^{(0,1/2)}(\sqrt{(p.v)^2-m^2}) \over \sqrt{p.v+m}\sqrt{p.v'+m}} 
\label{105-69e}
\eeq

\noindent One obtains, for the scalar product (\ref{105-66e}) :
$$<\psi^{(0,1/2,v',\epsilon ')} |\psi^{(0,1/2,v,\epsilon)} >\ = \overline{\epsilon '}\left[mA(w) + B(w){/\hskip - 2 truemm v} + C(w){/\hskip - 2 truemm v}'\right]\epsilon$$ 
\beq
= \left[mA(w) + B(w) + C(w)\right]\ \overline{\epsilon '}\epsilon
\label{105-70e}
\eeq

\noindent On the other hand, multiplying (\ref{105-68e}) by $v_\mu$ or $v'_\mu$ one can isolate the functions $B(w)$ and $C(w)$ and finally one gets 
$$\xi(w) = {1 \over w+1}\ \int {d^3\vec p \over p^0}\ \varphi^{(0,1/2)}(\sqrt{(p.v')^2-m^2})^* \varphi^{(0,1/2)}(\sqrt{(p.v)^2-m^2})$$
\beq
\label{105-71e}
\times\ {p.(v+v')+m(w+1) \over \sqrt{(p.v+m)(p.v'+m)}} 
\eeq

\noindent i.e. we find expression (\ref{105-1e}).\par

\vskip 5 truemm
\underline{Case $ j = 1/2, L = 0 \to j = 1/2, L = 1$} \par
\vskip 5 truemm

In this case the following invariant is involved :
\beq
\label{105-72e}
<\psi^{(1,1/2,v',\epsilon ')} |\psi^{(0,1/2,v,\epsilon)} >\ = \zeta(w)\ \overline{\epsilon '}\gamma_5\epsilon \qquad \qquad \qquad \ \ (\zeta(w) = 2\tau_{1/2}(w)) \qquad
\eeq

\noindent where we quote the two notations current in  the literature.\par
From (\ref{105-63e}), using the expressions for the $L = 1$ states  
\beq
\label{105-73e}
\psi^{(1,1/2,v,\epsilon)}(p) = - {1 \over \sqrt{2m(p.v-m)}} \gamma_5\ ({/\hskip - 2 truemm p}-m)\ \epsilon\ \varphi^{(1,1/2)}(\sqrt{(p.v)^2-m^2})
\eeq

\noindent and computing the scalar product (\ref{105-72e}), the calculation is very similar as for the ground state IW function and we obtain, after some algebra :
$$\zeta(w) = - {1 \over w-1}\ \int {d^3\vec p \over p^0}\ \varphi^{(1,1/2)}(\sqrt{(p.v')^2-m^2})^*\ \varphi^{(0,1/2)}(\sqrt{(p.v)^2-m^2})$$
\beq
\label{105-74e}
\times\ {1 \over \sqrt{(p.v+m)(p.v'+m)}}\ {1 \over \sqrt{(p.v')^2-m^2}}\ [(p.v')+m][(p.v')-(p.v)+m(w-1)]  
\eeq

\vskip 5 truemm
\underline{Case $ j = 1/2, L = 0 \to j = 3/2, L = 1$ } \par
\vskip 5 truemm

The following invariant is involved :
\beq
\label{105-75e}
<\psi^{(1,3/2,v',\epsilon')} |\psi^{(0,1/2,v,\epsilon)} >\ = \tau(w)\ (\overline{\epsilon '}.v) \epsilon \qquad \qquad \qquad \ (\tau(w) = \sqrt{3}\tau_{3/2}(w)) \ \ 
\eeq

\noindent where we quote the two notations used in the literature.\par

From (\ref{105-62e}) one gets
\beq
\label{105-76e}
\psi^{(1,3/2,v,\epsilon)}(p) = - {\sqrt{3} \over \sqrt{2m(p.v+m)}} ({/\hskip - 2 truemm p}+m)\ \epsilon.p\ {\varphi^{(1,3/2)}(\sqrt{(p.v)^2-m^2}) \over \sqrt{(p.v)^2-m^2}} 
\eeq

The scalar product (\ref{105-75e}) writes :
\beq
\label{105-77e}
<\psi^{(1,3/2,v',\epsilon ')} |\psi^{(0,1/2,v,\epsilon)} >\ = \int {d^3\vec p \over p^0}\ p_\mu \overline{\epsilon '}^\mu ({/\hskip - 2 truemm p}+m)\epsilon\ F(p,v,v')
\eeq

\noindent where now
$$F(p,v,v') = - {\sqrt{3} \over \sqrt{p.v+m}\sqrt{p.v'+m}}$$
\beq
\times {\varphi^{(1,3/2)}(\sqrt{(p.v')^2-m^2})^* \over  \sqrt{(p.v')^2-m^2}}\ \varphi^{(0,1/2)}(\sqrt{(p.v)^2-m^2}) 
\label{105-78e}
\eeq

\noindent We have now to compute the integrals
\beq
\int {d^3\vec p \over p^0}\ p_\mu\ F(p,v,v') = A(w) v_\mu + B(w) v'_\mu
\label{105-79e}
\eeq
\beq
\int {d^3\vec p \over p^0}\ p_\mu p_\nu\ F(p,v,v') = C(w) v_\mu v_\nu + D(w) (v_\mu v'_\nu + v'_\mu v_\nu) + E(w) v'_\mu v'_\nu + G(w) g_{\mu \nu}
\label{105-80e}
\eeq

\noindent Using the conditions (\ref{105-56bise})(c,d,e) one sees from (\ref{105-75e}) that the IW function is given in terms of only three functions
\beq
\tau(w) = C(w) + D(w) + m A(w)
\label{105-81e}
\eeq

\noindent Saturating the index $\mu$ in (\ref{105-79e}) with $v^\mu$ and $v'^\mu$ one finds the equations
$$A(w) + wB(w) = \int {d^3\vec p \over p^0}\ (v.p)\ F(p,v,v')$$
\beq
wA(w) + B(w) = \int {d^3\vec p \over p^0}\ (v'.p)\ F(p,v,v')
\label{105-82e}
\eeq

\noindent and saturating  the indices $\mu, \nu$ in (\ref{105-80e}) with the tensors $v^\mu v^\nu, v^\mu v'^\nu, ...\ g^{\mu \nu}$ one gets the set of linear equations
$$C(w) + 2wD(w) + w^2 E(w) + G(w) = \int {d^3\vec p \over p^0}\ (v.p)^2\ F(p,v,v')$$
$$wC(w) + (w^2+1)D(w) + wE(w) + wG(w) = \int {d^3\vec p \over p^0}\ (v.p)(v'.p)\ F(p,v,v')$$
$$w^2 C(w) + 2wD(w) + E(w) + G(w) = \int {d^3\vec p \over p^0}\ (v'.p)^2\ F(p,v,v')$$
\beq
C(w) + 2wD(w) + E(w) + 4G(w) = \int {d^3\vec p \over p^0}\ m^2\ F(p,v,v')
\label{105-83e}
\eeq

\noindent Equations (\ref{105-82e})(\ref{105-83e}) allow to compute the different functions $A(w)$,... $G(w)$. From these functions and (\ref{105-81e}) one finally gets :
$$\tau(w) = - {\sqrt{3} \over 2(w-1)(w+1)^2}\ \int {d^3\vec p \over p^0}\ \varphi^{(1,3/2)}(\sqrt{(p.v')^2-m^2})^*\ \varphi^{(0,1/2)}(\sqrt{(p.v)^2+m^2})$$
$$\times\ {1 \over \sqrt{(p.v+m)(p.v'+m)}}\ {1 \over  \sqrt{(p.v')^2-m^2}}\ [-3(p.v)^2+(2w-1)(p.v')^2$$
\beq
\label{105-84e}
+\ 2(2w-1)(p.v)(p.v')+2(w+1)(w(p.v')-(p.v))m-(w^2-1)m^2] 
\eeq

Taking into account differences in definition and normalization conventions, the expressions (\ref{105-74e}) and (\ref{105-84e}) are the same as found in the previous papers \cite{13bisr,MLOPR-1}.

\section{Conclusions}

We have applied the Lorentz group method to study the Isgur-Wise function in the case of mesons $B \to D^{(*)} \ell \nu$ where the light quark has $j = {1 \over 2}$. We recover the constraints obtained previously using the Bjorken-Uraltsev sum rule method, plus a number of other results.\par

In particular, we have obtained an integral representation for the IW function in terms of elementary functions and a positive measure. We have inverted this representation, expressing the measure in terms of the IW function. This has allowed us to test whether a given ansatz of the IW function satisfies the Lorentz or, equivalently, the generalized Bjorken-Uraltsev SR constraints.\par

We have compared a number of phenomenological shapes for the Isgur-Wise function with the obtained theoretical constraints. This has provided explicit illustrations of the method in a rather complete way. The different criteria based on the Lorentz group, i.e. lower limits on the derivatives at zero recoil, positivity of the measure in the inversion formula for the IW function and the upper bound for the whole IW function, have been illustrated by using different models of the IW function.\par

We have studied a number of models proposed in the literature : exponential shape, "dipole" form, Kiselev ansatz, Bauer-Stech-Wirbel model, relativistic harmonic oscillator, QCD Sum Rules, Bakamjian-Thomas relativistic quark model, etc. We have shown that the "dipole", the BSW model and the BT model satisfy the theoretical constraints.\par

The case of the QCDSR result is particularly interesting because of its link to general principles. In the limit in which the condensates are disregarded, the predicted dipole shape satisfies all the constraints. However, switching on the condensates spoils this nice feature. Of course, one can argue that the OPE has been limited to the lower dimension condensates. Our results show the interesting feature that in this framework one could obtain incorrect results by keeping only the lowest dimension operators. Our study in the heavy quark limit
does not take into account the radiative corrections, in consistency with the considered theoretical hypothesis - factorization between the heavy quark matrix element and the light could overlap - in which the methods of the present paper can hold.\par

We have studied in detail the Bakamjian-Thomas relativistic quark model applied to mesons in the heavy quark limit. To this aim we have described the Lorentz group representation that underlies the model. We formulate the form of the wave functions of the light cloud for all quantum numbers, and provide the formalism to obtain the IW functions by scalar products of these states. Consistently, the elastic IW function in this model satisfies all the Lorentz group criteria, and this feature holds for any explicit form of the Hamiltonian describing the meson spectrum at rest. Completeness in the Hilbert space implies the strong result that the full set fo Bjorken-like heavy quark limit sum rules is automatically satisfied in the BT model at infinite mass.\par

In conclusion, using a method based on the Lorentz group, completely equivalent to the one of the generalized Bjorken-Uraltsev sum rules, we have obtained in this paper strong constraints on the Isgur-Wise function for the ground state mesons.

\section*{Acknowledgement} \hspace*{\parindent} 
We acknowledge discussions with Fr\' ed\' eric Jugeau in the early stages of this work. 

\end{document}